\documentclass[article,12pt]{amsart}

\usepackage{amsmath}
\usepackage{amsfonts}
\usepackage{amssymb}
\usepackage{graphicx}
\usepackage{enumerate}
\usepackage{color}
\usepackage{array}
\usepackage{bbm}
\usepackage{fullpage}

\allowdisplaybreaks

\numberwithin{equation}{section}

\newcommand{\dE}{\mathbb{E}}
\newcommand{\dG}{\mathbb{G}}
\newcommand{\dN}{\mathbb{N}}
\newcommand{\dP}{\mathbb{P}}

\newcommand{\dT}{\mathbb{T}}

\newcommand{\cG}{\mathcal{G}}
\newcommand{\cI}{\mathcal{I}}
\newcommand{\cL}{\mathcal{L}}
\newcommand{\cP}{\mathcal{P}}
\newcommand{\cR}{\mathcal{R}}
\newcommand{\cS}{\mathcal{S}}
\newcommand{\cT}{\mathcal{T}}

\newcommand{\de}{\mathrm{e}}

\newcommand{\ind}{\mathbbm{1}}
\newcommand{\card}{\textnormal{Card}}
\newcommand{\hsp}{\hspace{0.5cm}}
\newcommand{\wh}{\widehat}
\newcommand{\ex}{\smallskip \noindent \textit{Example. }}
\newcommand{\exs}{\smallskip \noindent \textit{Examples. }}

\renewcommand{\leq}{\leqslant}
\renewcommand{\geq}{\geqslant}


\begin{document}

\title[Probabilistic reconstruction of genealogies]
{Probabilistic reconstruction of genealogies for polyploid plant species
\vspace{2ex}}
\author[F. Pro\"ia]{Pro\"ia Fr\'ed\'eric}
\author[F. Panloup]{Panloup Fabien}
\author[C. Trabelsi]{Trabelsi Chiraz}
\address{Laboratoire angevin de recherche en math\'ematiques, LAREMA, UMR 6093, CNRS, UNIV Angers, SFR MathSTIC, 2 Bd Lavoisier, 49045 Angers Cedex 01, France.}
\email{frederic.proia@univ-angers.fr}
\email{fabien.panloup@univ-angers.fr}
\email{chiraz.trabelsi@univ-angers.fr}
\author[J. Clotault]{Clotault J\'er\'emy}
\address{IRHS, Agrocampus-Ouest, INRA, Universit\'e d'Angers, SFR 4207 QuaSaV, 49071, Beaucouz\'e, France.}
\email{jeremy.clotault@univ-angers.fr}


\begin{abstract}
A probabilistic reconstruction of genealogies in a polyploid population (from 2x to 4x) is investigated, by considering genetic data analyzed as the probability of allele presence in a given genotype. Based on the likelihood of all possible crossbreeding patterns, our model enables us to infer and to quantify the whole potential genealogies in the population. We explain in particular how to deal with the uncertain allelic multiplicity that may occur with polyploids. Then we build an \textit{ad hoc} penalized likelihood to compare genealogies and to decide whether a particular individual brings sufficient information to be included in the taken genealogy. This decision criterion enables us in a next part to suggest a greedy algorithm in order to explore missing links and to rebuild some connections in the genealogies, retrospectively. As a by-product, we also give a way to infer the individuals that may have been favored by breeders over the years. In the last part we highlight the results given by our model and our algorithm, firstly on a simulated population and then on a real population of rose bushes. Most of the methodology relies on the maximum likelihood principle and on graph theory.
\end{abstract}

\maketitle

\section{Introduction}
\label{SecIntro}

\subsection{Motivations}
\label{SecIntroMotiv}

Pedigrees depict the genealogical relationships between individuals of a given population. They can be built thanks to mating knowledge or they can be inferred from molecular markers. The identification of pedigrees allows a broad variety of applications: genealogy identification, like in grapevine \cite{Lacombe2013}, improvement of conservation programs for endangered species \cite{Lucena-Perez2018}, inference of statistics used in quantitative and population genetics like heritability or population effective size \cite{Ackerman2017,Kong2015}, etc. Like for most population genetics analyses, pedigree reconstruction methods and their implementation were firstly developed for diploid species (but see \cite{Wang2014}). Polyploids, \textit{i.e.} species with more than two alleles for a given locus, represent approximately 25\% of plant species \cite{Barker2016}, and among them a large number of cultivated species. Polyploidy in animals is more rare but some examples were described in insects, fishes, amphibians and reptiles \cite{Otto2000,Mable2011}.

\smallskip

Several strategies were used to reconstruct the genealogical relationships from molecular markers (reviewed in \cite{Jones2003}). Exclusion methods eliminate potential parents which do not show at least one allele per locus shared with a putative offspring. If more than two parents are possible, categorical allocation methods allow identification of the most likely parents according to their probability to transmit alleles shared with the potential progeny. Parental reconstruction methods use full- or half-siblings in order to identify the most likely parents. By comparison, sibling reconstruction methods add a preliminary step of inference of siblings when they are unknown. In this paper, the objective is to adapt and to extend the approach of \cite{Chaumont2017}, namely to determine for each individual the most likely couple of parents amongst all older individuals, so as to build some family trees in polyploid plant species. Our study certainly intends to be applied on real genetic datasets, in particular the main practical motivation is to find some retrospective links in a population of rose bushes that will now be described.

\smallskip

The empirical dataset used in the last section of this article was obtained on cultivated roses bred mainly during the nineteenth century (\textit{Rosa} sp.). Rose breeding activities were particularly abundant during the nineteenth century and were very documented. As an example, breeding year is known for a majority of roses from this period. However, the genealogical relationships described in archives are highly hypothetical, due to the lack of control of artificial hybridization until the end of the nineteenth century. Among the approximately 200 species of the genus \textit{Rosa}, ploidy level varies between 2x and 10x \cite{Jian2010}. Rose breeding activities from the nineteenth century involved interspecific crossings between diploid species and tetraploid species, with a small contribution of genotypes with higher ploidy like species from the \textit{Caninae} section (4x, 5x and 6x) \cite{Oghina-Pavie2015,Liorzou2016}. Cultivated roses bred during the nineteenth century can exhibit all ploidy values between 2x and 6x, even if 5x and 6x are rare \cite{Liorzou2016}. The mode of inheritance in these rose cultivars remains highly unknown. It is generally considered that modern tetraploid cultivated roses exhibit a tetrasomic inheritance (no preferred pairing among the set of four homologous chromosomes and creation of tetravalents during meiosis) \cite{Koning-Boucoiran2012}. But a mixture of disomic (preferred pairing of two bivalent pairs during meiosis according to their genomic similarity) and tetrasomic inheritance could be observed according to chromosomes and according to genotypes \cite{Bourke2017}. Triploid roses have played a major role in rose hybridizations. Like in other species, triploid roses exhibit a low fertility rate, due to irregular meiosis leading to aneuploidy \cite{maia1976cytotaxonomie}. However, even if the production of fertile gametes from triploids remains rare, these events were selected by breeders, especially as bridges between different ploidy levels. For example, \textit{Bourbon}, \textit{Hybrid China} and \textit{Hybrid Tea} rose groups were both obtained by a cross between a Chinese diploid cultivar and a European tetraploid cultivar. First cultivars from these groups were triploid \cite{gudin2000rose}. Triploids form both haploid and diploid gametes \cite{van2005interploidy}. Following the obtention of a variety by hybridization, it was then propagated vegetatively by cutting or grafting and often conserved in rose gardens. Therefore rose varieties can be considered as immortal and they could have been involved at different periods in rose pedigrees. As most of plants, roses are hermaphrodites and can therefore have been used as female or male on different hybridization events. Selfing rate in roses is very low mainly because of self-incompatibility (\cite{Raju2013} and J. Mouchotte, pers. comm.). These specific breeding behaviors are the cornerstone of our probabilistic model.

\smallskip

In a general way, the polyploidy of the population may give rise to complications in terms of multiplicity of the alleles, being only aware of their presence or absence: that will be one of our strategic challenges to deal with this lack of information, widely discussed throughout the manuscript. Whereas for diploids the presence or absence of alleles is sufficient -- for $\{ a \}$ and $\{ a,b \}$ undoubtedly correspond to $\{ a,a \}$ and $\{ a,b \}$ -- the observation of $\{ a,b \}$ for a tetraploid can correspond to $\{ a,a,a,b \}$, $\{ a,a,b,b \}$ or $\{ a,b,b,b \}$. Reading the presence or absence of alleles on electrophoregrams and interpreting theoretical ratios between peak intensities is an option to determine the number of copies of each allele \cite{Esselink2004}. Unfortunately, we will explain in good time the reasons why this strategy is not reliable in our context and we will introduce a way to deal with this allelic multiplicity through the intermediary of probabilities related to each configuration. Before getting to the heart of the matter, let us point out that the objective of this work is not to introduce a biological issue, but rather to build and justify the more realistic mathematical framework regarding the biological model of roses bred during the nineteenth century. This work is above all a methodological one.

\smallskip

The paper is organized as follows. In Section \ref{SecGen}, we present a probabilistic method in order to reconstruct genealogies for species with several ploidy levels, from 2x to 4x, by considering genetic data analyzed as the probability of allele presence in a given genotype. In particular, we compute the likelihood associated with all crossbreeding patterns and we explain how to build and quantify the whole possible genealogies of the population and how to treat the unknown allelic multiplicity. As a by-product we also give a way to find the individuals favored by breeders, retrospectively. Section \ref{SecMiss} treats the isolated individuals, more precisely, the missing links. Under some criteria, we suggest an algorithm computing virtual individuals to improve the genealogy. Whereas Sections \ref{SecGen} and \ref{SecMiss} are mainly theoretical, all our results will be tested in Section \ref{SecEmp}, both on a simulated population and on a rose bushes population. We conclude by highlighting some weaknesses of our methodology and by giving, in accordance, some trails for future studies.

\subsection{Preliminary considerations and notations}
\label{SecIntroTech}

In the whole paper, $\cP$ stands for the population of size $n = \card(\cP)$ and $m$ is the number of genes involved in the reconstruction process. Technically, $m$ corresponds to the number of signals on which we read the \textit{peaks}, expressing the set of alleles detected on each gene. We make the crucial hypothesis that signals are \textit{mutually independent}, which can be argued on a genetic as well as statistical point of view (genes are chosen for their absence of known interaction and a prior statistical treatment tends to decorrelate them by eliminating material-type influences). For an individual $e \in \cP$, we denote by $g_{s}(e)$ the \textit{genotype} of gene $s$, that is, the \textit{set of alleles} present for this gene, shortened in $g(e)$ when we deal with an unspecified gene (to be precise, we should in fact speak of \textit{multiset} since we may have multiple instances of the same allele in the genotype, however we shall not make these kind of distinctions). We also denote by $x(e) = \card(g(e)) \in \{ 2,3,4 \}$ the \textit{ploidy} of $e$, the number of sets of chromosomes in a cell. In addition, we assume that the birth dates are known and that no death occurs, which is consistent with the fact that the work is related to plant cultivars. We also assume that gametes are produced according to strict polysomic inheritance and we neglect double reduction.

\section{Likelihood of a genealogy}
\label{SecGen}

This section is the heart of the paper. Firstly we will describe the genetic patterns that we retain to cross the polyploid individuals, and we will discuss the probabilistic treatment of the allelic multiplicity that may appear for triploids and tetraploids. Thereafter, we will be in the position to estimate some retrospective links and to compute an \textit{ad hoc} penalized likelihood for the genealogy. Before anything else, let us begin with a formal description of what we mean by genealogy and likelihood. A genealogy is an element of the set
\begin{equation}
\label{SetOfGen}
\Upsilon(\cP) = \prod_{e\, \in\, \cP} \big\{ \dT(e)\, \cup\, (e,\, \varnothing) \big\} \hsp \text{where} \hsp \dT(e) = \bigcup_{s\, \in\, \cS(e)} (e,\, s)
\end{equation}
and where $\cS(e)$, as will be detailed in good time (see beginning of Subsection \ref{SecGenProb}), is the set of non-ordered pairs candidates to the genealogy of $e$. In concrete terms, an individual $e$ is associated with each couple of possible parents $\cS(e)$ together with $\varnothing$, to cover the case where $\dT(e) = \varnothing$, that is where no triplet offspring/couple of parents can be found in the population for $e$. Thus, a genealogy $\cT$ on $\cP = \{ e_1, \hdots, e_{n} \}$ takes the form of
\begin{equation}
\label{FormOfGen}
\cT = \big\{ (e_1, s_1),\, (e_2, s_2),\, \hdots,\, (e_{n}, s_{n}) \big\}
\end{equation}
in which $s_{i}$ is either an element of $\cS(e_{i})$, either $\varnothing$. It has clearly a structure of graph, as will be explained later. Now, looking at $\cT$ as the realization of a discrete random vector taking values in the set $\Upsilon(\cP)$, it naturally follows that the \textit{likelihood of a genealogy} is the probability that it has to be observed, in accordance with the statistical usual definition, given a model and a set of hypotheses that will be described in this section. It should also be noted that a maximum likelihood genealogy, as will be largely discussed later, is not an estimator in the statistical sense, but the value of $\cT \in \Upsilon(\cP)$ having the biggest probability, with respect to the model.

\subsection{Crossbreeding patterns}
\label{SecGenPat}

\begin{figure}[h!]
\includegraphics[height=5cm]{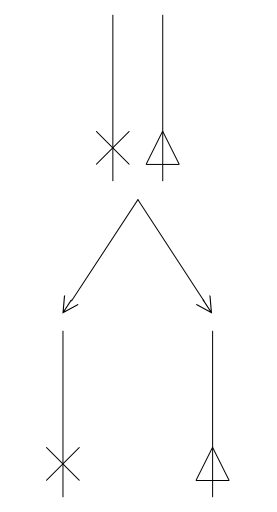}
\caption{Schematic representation of the gametes production (in the bottom) for a diploid cell (in the top). Symbols represent the alleles of a given gene on its chromosome (line).}
\label{FigSchGam2x}
\end{figure}

\begin{figure}[h!]
\includegraphics[height=5cm]{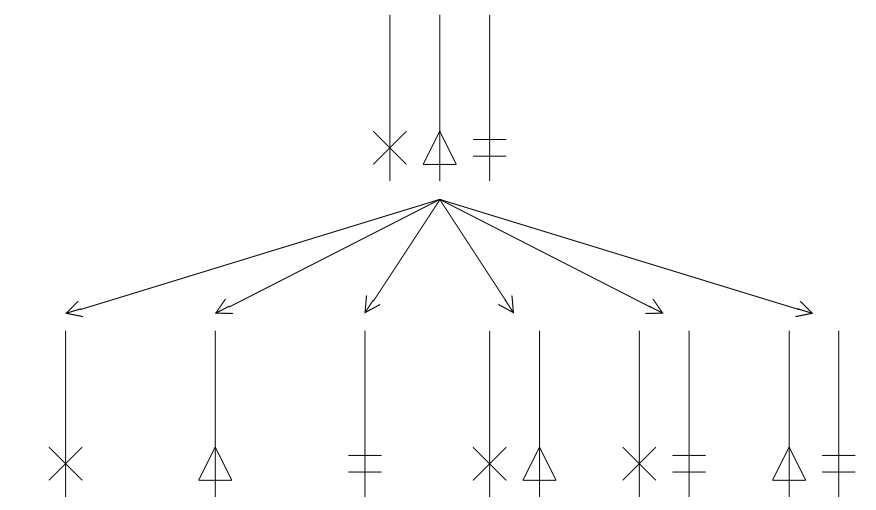}
\caption{Schematic representation of the gametes production (in the bottom) for a triploid cell (in the top). Symbols represent the alleles of a given gene on its chromosome (line).}
\label{FigSchGam3x}
\end{figure}

\begin{figure}[h!]
\includegraphics[height=5cm]{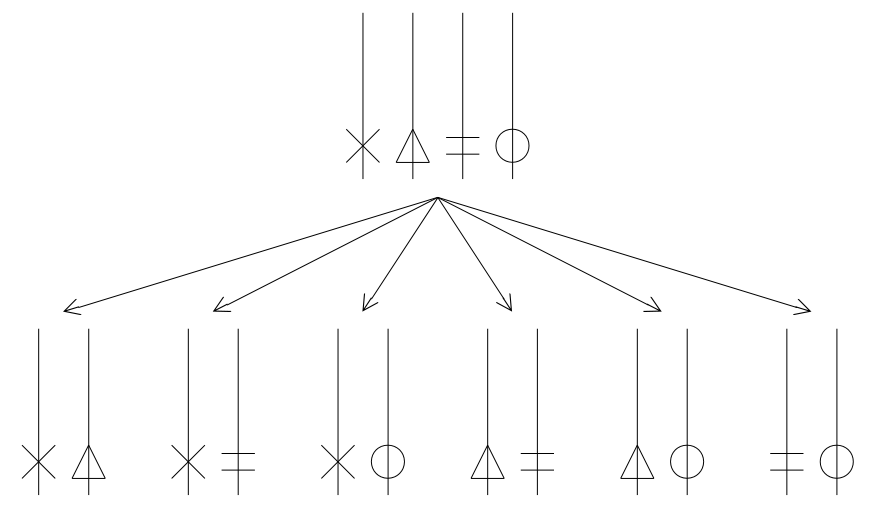}
\caption{Schematic representation of the gametes production (in the bottom) for a tetraploid cell (in the top). Symbols represent the alleles of a given gene on its chromosome (line).}
\label{FigSchGam4x}
\end{figure}

To simplify the combinatorial analysis, we use the following natural models. Diploids produce haploid gametes, genotype $\{ a, b \}$ leads to gametes $\{ a \}$ and $\{ b \}$ with probability $1$. Triploids produce haploid and diploid gametes, genotype $\{ a, b, c \}$ leads to gametes $\{ a \}$ and $\{ b, c \}$ with probability $\frac{1}{3}$, gametes $\{ b \}$ and $\{ a, c \}$ with probability $\frac{1}{3}$ and gametes $\{ c \}$ and $\{ a, b \}$ with probability $\frac{1}{3}$. Tetraploids produce diploid gametes, genotype $\{ a, b, c, d \}$ leads to gametes $\{ a, b \}$ and $\{ c, d \}$ with probability $\frac{1}{3}$, gametes $\{ a, c \}$ and $\{ b, d \}$ with probability $\frac{1}{3}$ and gametes $\{ a, d \}$ and $\{ b, c \}$ with probability $\frac{1}{3}$. In addition, each individual can either be male or female, the set of gametes is treated as an \textit{urn problem}. Crossing is made by choosing at random two gametes among all these possibilities, bringing them together to obtain the offspring's genotype. Figures \ref{FigSchGam2x}--\ref{FigSchGam3x}--\ref{FigSchGam4x} are schematic representations of the gametes production, indicated by arrows, of a parent cell.

Let $p_1$ and $p_2$ be two individuals having ploidies $x(p_1)$ and $x(p_2)$ with genotypes $g(p_1) = \{ a_1, \hdots, a_{x(p_1)} \}$ and $g(p_2) = \{ b_1, \hdots, b_{x(p_2)} \}$, respectively. In the sequel, $p_1$ and $p_2$ are the parents of the offspring $e$. The different ploidy levels lead to six patterns that we are now going to describe in detail.
\begin{enumerate}[(P$_1$)]
\item \label{Sch22} $x(p_1) = x(p_2) = 2$. Let $g(p_1) = \{ a_1, a_2 \}$ and $g(p_2) = \{ b_1, b_2 \}$. Then, $e$ has 4 potential diploid genotypes $g(e) = \{ a_{i}, b_{k} \}$, for $i,k \in \{ 1,2 \}$. Each one has probability $\frac{1}{4}$.
\item \label{Sch23} $x(p_1) = 2$ and $x(p_2) = 3$. Let $g(p_1) = \{ a_1, a_2 \}$ and $g(p_2) = \{ b_1, b_2, b_3 \}$. Then, $e$ has 6 potential diploid genotypes $g(e) = \{ a_{i}, b_{k} \}$, and 6 potential triploid genotypes $g(e) = \{ a_{i}, b_{k}, b_{\ell} \}$, for $i \in \{ 1,2 \}$ and $k,\ell \in \{ 1,2,3 \}$. Each one has probability $\frac{1}{12}$.
\item \label{Sch24} $x(p_1) = 2$ and $x(p_2) = 4$. Let $g(p_1) = \{ a_1, a_2 \}$ and $g(p_2) = \{ b_1, b_2, b_3, b_4 \}$. Then, $e$ has 12 potential triploid genotypes $g(e) = \{ a_{i}, b_{k}, b_{\ell} \}$, for $i \in \{ 1,2 \}$ and $k,\ell \in \{ 1,2,3,4 \}$. Each one has probability $\frac{1}{12}$.
\item \label{Sch33} $x(p_1) = x(p_2) = 3$. Let $g(p_1) = \{ a_1, a_2, a_3 \}$ and $g(p_2) = \{ b_1, b_2, b_3 \}$. Then, $e$ has 9 potential diploid genotypes $g(e) = \{ a_{i}, b_{k} \}$, 18 potential triploid genotypes $g(e) = \{ a_{i}, b_{k}, b_{\ell} \}$ or $g(e) = \{ a_{i}, a_{j}, b_{k} \}$, and 9 potential tetraploid genotypes $g(e) = \{ a_{i}, a_{j}, b_{k}, b_{\ell} \}$, for $i,j,k,\ell \in \{ 1,2,3 \}$. Each one has probability $\frac{1}{36}$.
\item \label{Sch34} $x(p_1) = 3$ and $x(p_2) = 4$. Let $g(p_1) = \{ a_1, a_2, a_3 \}$ and $g(p_2) = \{ b_1, b_2, b_3, b_4 \}$. Then, $e$ has 18 potential triploid  genotypes $g(e) = \{ a_{i}, b_{k}, b_{\ell} \}$, and 18 potential tetraploid genotypes $g(e) = \{ a_{i}, a_{j}, b_{k}, b_{\ell} \}$, for $i,j \in \{ 1,2,3 \}$ and $k,\ell \in \{ 1,2,3,4 \}$. Each one has probability $\frac{1}{36}$.
\item \label{Sch44} $x(p_1) = x(p_2) = 4$. Let $g(p_1) = \{ a_1, a_2, a_3, a_4 \}$ and $g(p_2) = \{ b_1, b_2, b_3, b_4 \}$. Then, $e$ has 36 potential tetraploid genotypes $g(e) = \{ a_{i}, a_{j}, b_{k}, b_{\ell} \}$, for $i,j,k,\ell \in \{ 1,2,3,4 \}$. Each one has probability $\frac{1}{36}$.
\end{enumerate}

\smallskip

\begin{figure}[h]
\includegraphics[height=6cm]{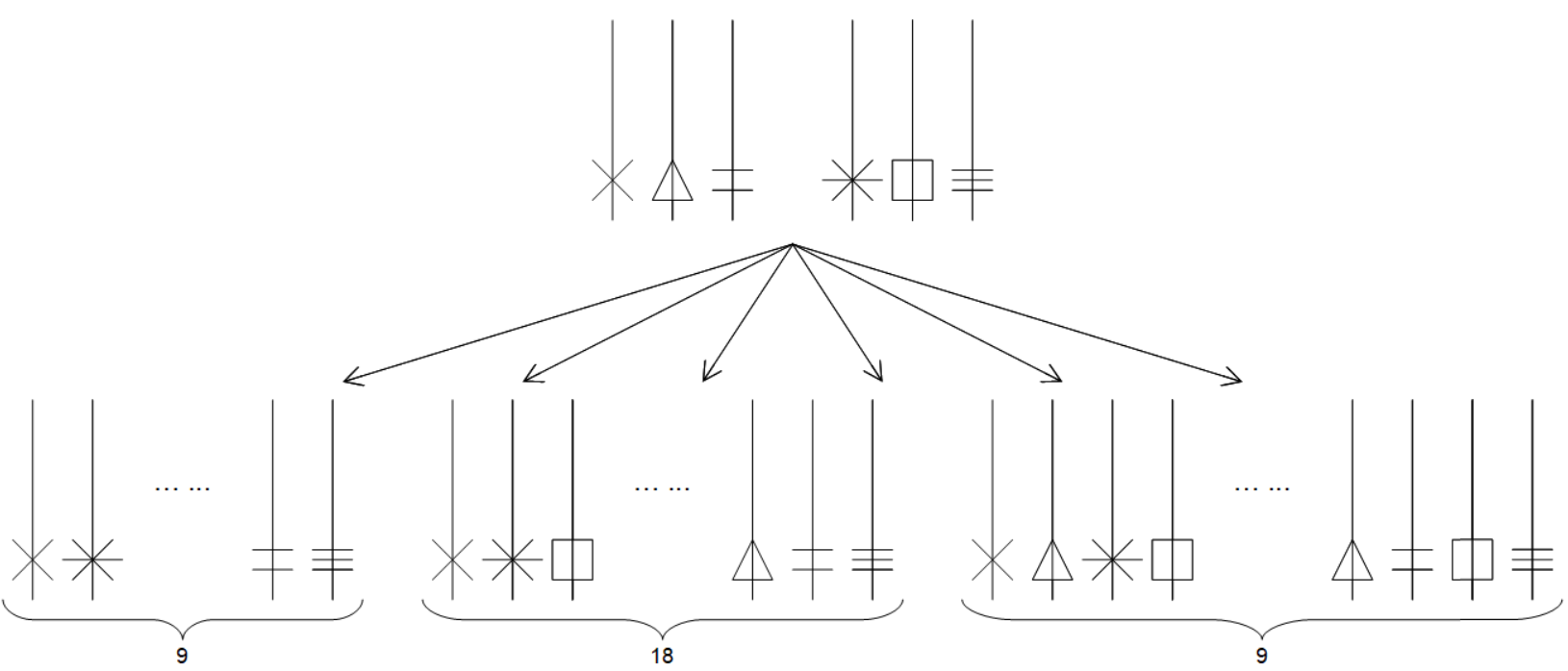}
\caption{Schematic representation of pattern (P$_{\ref{Sch33}}$) leading to $u=36$ potential offsprings including 9 diploids, 18 triploids and 9 tetraploids. Symbols represent the alleles of a given gene.}
\label{FigSch3x3x}
\end{figure}

To sum up, all diploid offsprings may come from patterns (P$_{\ref{Sch22}}$)--(P$_{\ref{Sch23}}$)--(P$_{\ref{Sch33}}$), all triploid offsprings from patterns (P$_{\ref{Sch23}}$)--(P$_{\ref{Sch24}}$)--(P$_{\ref{Sch33}}$)--(P$_{\ref{Sch34}}$) and all tetraploid offsprings from patterns (P$_{\ref{Sch33}}$)--(P$_{\ref{Sch34}}$)--(P$_{\ref{Sch44}}$). One can remark that the trickiest case is probably (P$_{\ref{Sch33}}$) since three different ploidies can be generated by crossing triploids, Figure \ref{FigSch3x3x} gives a streamlined representation of it.
Now let $\{ (p_1, p_2) \mapsto e \}$ be the event through which the pair $(p_1, p_2)$ conceives $e$, let $u$ denote the maximum number of different genotypes generated by the pattern ($u = 4$, $u = 12$ or $u = 36$) corresponding to the ploidy of $p_1$ and $p_2$, and let $e_1, \hdots, e_{u}$ name the potential offsprings of the cross. Our hypotheses show that, conditionally on the knowledge of the genotypes of the parents, each offspring is drawn through a uniform distribution. So, we set
\begin{equation}
\label{ProbaSch}
\dP( \{ (p_1, p_2) \mapsto e \}\, \vert\, \{ g(p_1), g(p_2), g(e) \} ) = \frac{1}{u} \sum_{r=1}^{u} \ind_{ \{ e_{r}\, =\, e \} }
\end{equation}
where the genetic equality $e_{r} = e$ means that $g(e_{r})$ and $g(e)$ coincide in a sense that we have to define. Specifically, we consider that $e_{r} = e$ once
\begin{equation}
\label{EqGen}
g(e_{r}) = g(e) \hsp \text{and hence} \hsp x(e_{r}) = x(e)
\end{equation}
which in this case amounts to say that $e_{r}$ and $e$ have the same ploidy and the same set of alleles (we remind that $x = \card(g)$). However, it is important to highlight that \eqref{EqGen} is only relevant from theoretical perspectives or on simulated data. We will see in Section \ref{SecEmpRos} that real genotypes result from a calibration of the equipment and some rounded values to be interpreted as \textit{base pairs}. Therefore,
\begin{equation}
\label{EqGenArr}
x(e_{r}) = x(e) \hsp \text{and} \hsp \Vert g^{*}(e_{r}) - g^{*}(e) \Vert_{\infty} \leq 1
\end{equation}
where $g^{*}$ stands for an ascending sorted vector containing the elements of $g$, should be an appropriate comparison on such data. Indeed, this criterion allows an offset of $\pm 1$ base pairs for two corresponding alleles.

\exs To illustrate this calculation method, let us consider $g(p_1) = \{ a, a \}$ and $g(p_2) = \{ a, b \}$. Then $u=4$, the potential offsprings have genotypes $g(e_1) = g(e_3) = \{ a, a \}$ and $g(e_2) = g(e_4) = \{ a, b \}$. For $g(e) = \{ a, a \}$ or $g(e) = \{ a, b \}$, formula \eqref{ProbaSch} gives probability $\frac{1}{2}$. It also gives probability 0 for all other genotypes. In the more intricate case where $g(p_1) = \{ a, a, b, c \}$ and $g(p_2) = \{ a, c, c \}$, then $u=36$ and among the potential offsprings, 5 will have genotype $g(e) = \{ a, b, c \}$. Formula \eqref{ProbaSch} gives probability $\frac{5}{36}$ for such a triploid offspring.

\subsection{Allelic multiplicity}
\label{SecGenMul}

For an individual $e \in \cP$, the set $g(e)$ is the \textit{true} genotype. However in our experimental studies, we only observe a \textit{partial} genotype $\wh{g}(e) \subset g(e)$ containing the distinct alleles -- a set of peaks on the signal. Taking advantage of the ploidy $x(e)$, one is able to infer all possible $g(e)$ from $\wh{g}(e)$. Explicitly, we use the following connections, where $\pi$ names a probability of multiplicity in a generic way.
\begin{enumerate}[(C$_1$)]
\item $\wh{g}(e) = \{ a \}$ and $x(e) = 2$ leads to $g(e) = \{ a, a \}$ with probability 1.
\item $\wh{g}(e) = \{ a, b \}$ and $x(e) = 2$ leads to $g(e) = \{ a, b \}$ with probability 1.
\item $\wh{g}(e) = \{ a \}$ and $x(e) = 3$ leads to $g(e) = \{ a, a, a \}$ with probability 1.
\item \label{Dos32} $\wh{g}(e) = \{ a, b \}$ and $x(e) = 3$ leads to $g(e) = \{ a, a, b \}$ with probability $\pi_{21}$ and to $g(e) = \{ a, b, b \}$ with probability $\pi_{12}$. We set $\pi_{21} + \pi_{12} = 1$.
\item $\wh{g}(e) = \{ a, b, c \}$ and $x(e) = 3$ leads to $g(e) = \{ a, b, c \}$ with probability 1.
\item $\wh{g}(e) = \{ a \}$ and $x(e) = 4$ leads to $g(e) = \{ a, a, a, a \}$ with probability 1.
\item \label{Dos42} $\wh{g}(e) = \{ a, b \}$ and $x(e) = 4$ leads to $g(e) = \{ a, a, a, b \}$ with probability $\pi_{31}$, $g(e) = \{ a, a, b, b \}$ with probability $\pi_{22}$ and $g(e) = \{ a, b, b, b \}$ with probability $\pi_{13}$. We set $\pi_{31} + \pi_{22} + \pi_{13} = 1$.
\item \label{Dos43} $\wh{g}(e) = \{ a, b, c \}$ and $x(e) = 4$ leads to $g(e) = \{ a, a, b, c \}$ with probability $\pi_{211}$, $g(e) = \{ a, b, b, c \}$ with probability $\pi_{121}$ and $g(e) = \{ a, b, c, c \}$ with probability $\pi_{112}$. We set $\pi_{211} + \pi_{121} + \pi_{112} = 1$.
\item $\wh{g}(e) = \{ a, b, c, d \}$ and $x(e) = 4$ leads to $g(e) = \{ a, b, c, d \}$ with probability 1.
\end{enumerate}

\smallskip

Instead of selecting a genotype for $e$ when several are conceivable, that is, for combinations (C$_{\ref{Dos32}}$)--(C$_{\ref{Dos42}}$)--(C$_{\ref{Dos43}}$), the model that we introduce in the next section takes account of all possibilities weighted by their related probabilities. In fact, our model enables us to choose if necessary $\pi = \pi^{(s)}$ gene by gene or, equivalently, signal by signal, to consider the different interpretations of the relative amplitude of the peaks on each signal, for material reasons. We will describe it in more details in the beginning of Section \ref{SecEmpRos}.

\subsection{Probability of a genealogical link}
\label{SecGenProb}

For any individual $e \in \cP$, as it has been outlined in the introduction of the section, let $\cS(e) \subset \cP^{\, 2}$ be the \textit{compatible} subpopulation, that is, the set of non-ordered pairs $(p_1, p_2)$ with $p_1 \neq p_2$ (excluding selfing) genetically and chronologically candidates to the genealogy of $e$. It is worth noting that the only chronological constraint is obviously to consider that birth dates of descendants cannot be prior to the ones of their parents. In particular, the probabilities of ancestry are considered as \textit{time-invariant}: any individual has the same probability of being a parent, regardless of its birth date, excluding \textit{de facto} any generational model like Galton-Watson trees. This point of view is specific to plant species, and would clearly be irrelevant for animal populations. Whether the individual was obtained during the decade preceding the birth date of the offspring, or several centuries ago, because of the immortality and constant fertility given by a vegetative propagation, we assume that the probability of ancestry is the same. Our objective is to build a probability measure on $\cS(e)\, \cup\, \{ \varnothing \}$ quantifying the whole possible genealogical links of $e$, the element $\varnothing$ being added to cover the case where no parents can be found in the population. The hypothesis of mutual independence of the signals allows us to work on each signal and to multiply the results. Let
\begin{equation}
\label{ProbaIndiv}
\delta(e, p_1, p_2) = \prod_{s=1}^{m} \sum_{G\, \in\, \cG_{s}} \dP( \{ (p_1, p_2) \mapsto e \}\, \vert\, G )\, \dP(G)
\end{equation}
where $\cG_{s}$ is the set of all possible genotypes on signal $s$ for the triplet $(e, p_1, p_2)$. In the best-case scenario, $\card(\cG_{s}) = 1$ which means that $\wh{g}_{s}(p_1)$, $\wh{g}_{s}(p_2)$ and $\wh{g}_{s}(e)$ lead to no uncertain allelic multiplicity, and thus $\dP(G) = 1$. At worst, $\card(\cG_{s}) = 27$ meaning that $\wh{g}_{s}(p_1)$, $\wh{g}_{s}(p_2)$ and $\wh{g}_{s}(e)$ are in the situation (C$_{\ref{Dos42}}$) or (C$_{\ref{Dos43}}$), and $\dP(G)$ is the product of the related probabilities.

\ex Suppose that $x(p_1) = 3$, $x(p_2) = 4$, $x(e) = 4$ and that, on a particular signal $s$, we observe $\wh{g}_{s}(p_1) = \{ a, b \}$, $\wh{g}_{s}(p_2) = \{ a, c, d \}$ and $\wh{g}_{s}(e) = \{ a, d \}$. Then, $\card(\cG_{s}) = 18$. Indeed, we build $\cG_{s}$ by combining $\{ a, a, b \}$ and $\{ a, b, b \}$ for $p_1$, $\{ a, a, c, d \}$, $\{ a, c, c, d \}$ and $\{ a, c, d, d \}$ for $p_2$, and $\{ a, a, a, d \}$, $\{ a, a, d, d \}$ and $\{ a, d, d, d \}$ for $e$. For the first combination we have $\dP(G) = \pi^{(s)}_{21}\, \pi^{(s)}_{211}\, \pi^{(s)}_{31}$, for the second one $\dP(G) = \pi^{(s)}_{21}\, \pi^{(s)}_{211}\, \pi^{(s)}_{22}$, and so on.

\smallskip

It only remains to renormalize. Explicitly, with
\begin{equation}
\label{ProbaIndivSum}
\Delta(e) = \sum_{(p_1, p_2)\, \in\, \cS(e)} \delta(e, p_1, p_2)
\end{equation}
where $\delta(e, p_1, p_2)$ is given in \eqref{ProbaIndiv}, let
\begin{equation}
\label{ProbaIndivRenorm}
\forall\, (p_1, p_2) \in \cS(e), \hsp \nu_{e}((p_1, p_2)) = \left\{
\begin{array}{ll}
\frac{\delta(e, p_1, p_2)}{\Delta(e)} & \mbox{if } \Delta(e) > 0 \\
0 & \mbox{otherwise}
\end{array}
\right.
\end{equation}
and fix $\nu_{e}(\varnothing) = 1$ as soon as $\Delta(e) = 0$, and $\nu_{e}(\varnothing) = 0$ otherwise. Then clearly, $\nu_{e} : \cS(e)\, \cup\, \{ \varnothing \} \rightarrow [0,1]$ is a probability measure that can be applied to look for the whole genealogy of $e \in \cP$. To build the \textit{most likely genealogy}, we must pick
\begin{equation}
\label{ProbaIndivMax}
c^{\, *}(e) = \underset{c\, \in\, \cS(e)\, \cup\, \{ \varnothing \}}{\arg \max} ~ \nu_{e}(c).
\end{equation}
To be precise, $c^{\, *}(e)$ defined as above is not necessarily unique, in such case we arbitrarily pick one optimum at random. We will see in the sequel that choosing a genealogical link amongst others is not necessarily relevant, hence we also consider
\begin{equation}
\label{ProbaIndivFull}
G(e) = \left\{ c\, \in\, \cS(e)\, \cup\, \{ \varnothing \} ~ \vert ~ \nu_{e}(c) > 0 \right\}
\end{equation}
which represents the whole potential genealogical links of $e$ in our population $\cP$.

\subsection{A retrospective family tree}
\label{SecGenTree}

Now the objective is to compute $G(e)$ -- and thus $c^{\, *}(e)$ -- for all $e \in \cP$. In the framework of this study, a \textit{family tree} $\cT$ of the population $\cP$ is a set of triplets $(e, p_1, p_2)$ having probabilities $\nu_{e}((p_1, p_2)) > 0$, on the basis of $m$ genes, such that there is at most one triplet $(e, p_1, p_2)$ for any individual $e$, interpretable as the realization of the event $\{ (p_1, p_2) \mapsto e \}$, taking up the notation of the previous sections. We also require that a triplet $(e, p_1, p_2)$ is assigned to the node $e$ of the family tree as soon as $c^{\, *}(e) \neq \varnothing$, that is as soon as there exists at least one potential genealogical link for $e$. To make the connection with our formal introduction, a family tree $\cT$ completed by $(e,\varnothing)$ for each $e$ such that $c^{\, *}(e) = \varnothing$ is merely a genealogy as it is defined in \eqref{FormOfGen}. In an equivalent way, we build a \textit{graph} in which each individual is a vertex and each genealogical link is a couple of arcs (from the parents to the offspring). Note that the chronological constraint applied on $\cS(e)$ is sufficient to ensure that no cycle is present in the graph. The methods and algorithms that follow will be tested and applied in Section \ref{SecEmp}.

\subsubsection{Most likely trees} Combining all options of $G(e)$ for each $e \in \cP$ gives an exhaustive set of trees, all potential genealogies of the population that we will denote as $\dG(\cP)$ in \eqref{WholeGen}. However, on large datasets, this can be difficult due to the exponential growth of the combinations. Thus we look for criteria of selection, and first we define the log-likelihood of a family tree $\cT$ as follows,
\begin{equation}
\label{LogLikGen}
\ell(\cT) = \sum_{(e, p_1, p_2)\, \in\, \cT} \ln \nu_{e}((p_1, p_2)).
\end{equation}
Note that this expression corresponds to the likelihood of a genealogy as we have defined it beforehand, under the crucial hypothesis that each triplet offspring/couple of parents is independent of any other, which once again is specific to plant species. Clearly $\cP$ can be divided into $\cL = \{ e \in \cP\, \vert\, c^{\, *}(e) \neq \varnothing \}$ and $\cI = \{ e \in \cP\, \vert\, c^{\, *}(e) = \varnothing \}$, respectively the individuals having potential ancestors in the population, present as nodes in all family trees built according to our constraints, and the ones for which we have not been able to find any genealogical link, that we will describe as \textit{isolated}. Our model guarantees that maximizing $\ell(\cT)$ amounts to locally maximizing the log-probability of each link. To sum up,
\begin{equation}
\label{LogLikGenMax}
\max_{\cT\, \in\, \Upsilon(\cP)} ~ \ell(\cT) = \sum_{e\, \in\, \cL} \ln \nu_{e}(c^{\, *}(e))
\end{equation}
and this upper bound is reached by the tree $\cT^{\, *}$ built on all $e \in \cL$ associated with the pairs $c^{\, *}(e)$. We shall note that formula \eqref{LogLikGenMax} does not necessarily highlight a unique family tree, for some pairs $(p_1, p_2)$ may have the same probability of producing $e$. In this case, the maximization problem has more than one solution.

\subsubsection{Number of offsprings} Suppose now that the population is small enough to be able to compute
\begin{equation}
\label{WholeGen}
\dG(\cP) = \prod_{e\, \in\, \cP} G(e)
\end{equation}
where $G(e)$ is given in \eqref{ProbaIndivFull}. Namely, $\dG(\cP)$ contains the exhaustive set of potential genealogies of the population. Due to the combination of the options of all $G(e)$, $\card(\dG(\cP))$ may be very large. In fact such a Cartesian product is only conceptual, but quickly intractable for practical purposes leading to combinatorial explosions. Therefore, a \textit{threshold probability} must be used to select the genealogies of $\dG(\cP)$. Concretely, we can replace the definition of $G(e)$ in \eqref{ProbaIndivFull} by the more stringent
\begin{equation}
\label{ProbaIndivFullThresh}
G(e) = \left\{ c\, \in\, \cS(e)\, \cup\, \{ \varnothing \} ~ \vert ~ \nu_{e}(c) > \pi_{\min} \right\}
\end{equation}
for a given choice of $0 \leq \pi_{\min} < 1$, and the construction of $\dG(\cP)$ accordingly. If we define $N(i)$ as a random variable counting the offsprings of $i \in \cP$, then it could be interesting to give an estimation of its probability distribution so as to infer, retrospectively, the individuals favored by breeders. Our model directly suggests to use
\begin{equation}
\label{EstProbaSelect}
\forall\, k \in \dN, \hsp \wh{\dP}(N(i) = k) = \sum_{g\, \in\, \dG(\cP)} w_{g}\, \ind_{\{ n_{g}(i)\, =\, k \}}
\end{equation}
where $n_{g}(i)$ is the number of offsprings of $i$ in the genealogy $g$ and $w_{g}$ is a \textit{weighting} of the genealogy that can naturally be defined as the ratio between its likelihood and the sum of all likelihoods, \textit{i.e.}
\begin{equation}
\label{WeightGen}
w_{g} = \frac{\de^{\, \ell(g)}}{L(\cP)} \hsp \text{with} \hsp L(\cP) = \sum_{h\, \in\, \dG(\cP)} \de^{\, \ell(h)}
\end{equation}
keeping the notation of \eqref{LogLikGen}. It follows that
\begin{equation}
\label{EstEspSelect}
\wh{\dE}[N(i)] = \sum_{g\, \in\, \dG(\cP)} w_{g}\, n_{g}(i)
\end{equation}
may be a useful tool to decide whether $i$ has been favored by breeders, by comparison with the global mean value and a classical \textit{outlier threshold}. This approach will be illustrated on the rose bushes population of Section \ref{SecEmpRos}.

\ex Consider a set of 4 genealogies of likelihood $0.8$, $0.6$, $0.1$ and $0.02$, among which an individual $i$ has $0$, $1$, $1$ and $2$ offsprings, respectively. Then we propose estimating $\wh{\dP}(N(i) = 0) \approx 0.526$, $\wh{\dP}(N(i) = 1) \approx 0.461$, $\wh{\dP}(N(i) = 2) \approx 0.013$ and $\wh{\dP}(N(i) > 2) = 0$. For this individual, $\wh{\dE}[N(i)] \approx 0.487$.

\smallskip

To look at pairwise relationships in the population, it can also be meaningful to build a \textit{genealogical graph} made of all possible (weighted) links. In such a graph, we are not interested in the triplets offspring/couple of parents, but only in the pairs offspring/parent. For all $(i,j) \in \cP^{\, 2}$ and the same weights as in \eqref{WeightGen}, consider
\begin{equation}
\label{GenGraph}
W_{i\, \rightarrow\, j} = \sum_{g\, \in\, \dG(\cP)} w_{g}\, \ind_{\{ (i\, \rightarrow\, j)\, \in\, g \}}
\end{equation}
where $\{ (i \rightarrow j) \in g \}$ means that $i$ is a parent of $j$ in the genealogy $g$. The directed and weighted graph built on $W_{i\, \rightarrow\, j}$ amounts to the superposition of all genealogies except that the viewpoint is different: edges are not considered in pairs, but each one has a role of its own. However it is worth noting that, according to this model, the outflow from an individual is precisely its averaged number of offsprings \eqref{EstEspSelect}. Thus, these two approaches are numerically equivalent but they differ from the interpretation.

\subsubsection{Comparison of trees} For a fixed population of size $n$, since each tree contains the same number of links, maximizing the likelihood \textit{via} \eqref{LogLikGen} seems a suitable criterion. However, it cannot be trusted to compare trees with a different number of links. To understand this, let $\cP_{i}= \cP \cup \{ i \}$ be the same population enhanced with a new individual, from the last generation, such that $\delta(i, p_1, p_2) > 0$ for at least two pairs $(p_1, p_2) \in \cS(i)$. Then, for these pairs we get $\ln \nu_{i}((p_1, p_2)) < 0$, implying that $\ell(\cT) > \ell(\cT_{i})$, where $\cT^{}$ and $\cT_{i}$ are the family trees maximizing the likelihood on $\cP$ and $\cP_{i}$, respectively. In other words, this criterion favors $\cT$ rather than $\cT_{i}$ whereas there exists a link between some individuals of $\cP$ and $i$. In order to overcome this negative impact, as soon as we have to compare family trees on two populations $\cP$ and $\cP_{i}$ such that $\cP_{i} = \cP \cup \{ i \}$, we suggest to consider a trade-off like
\begin{equation}
\label{LogLikPenGen}
\ell^{\, *}(\cT_{i}) = \ell(\cT_{i}) + \Psi(i)
\end{equation}
where $\ell(\cT_{i})$ is the log-likelihood given by \eqref{LogLikGen} of the genealogical tree $\cT_{i}$ on $\cP_{i}$ containing $i$, and $\Psi(i)$ is a measure of the \textit{interaction ability} of the new individual $i$ with $\cP$. Whence, to decide whether $i$ has to be added into the genealogy, it will be possible to compare $\ell^{\, *}(\cT_{i})$ and $\ell(\cT)$ for the most likely tree $\cT$ built on $\cP$, provided a suitable adjustment of $\Psi(i)$. In this way, we intend to compensate the mechanical decrease of the log-likelihood due to the accumulation of potential links including $i$. This penalization of the log-likelihood is a strategy similar to the well-known AIC and BIC criteria. In the next section, when looking for missing individuals that could improve the family tree, we will see how to give a suitable explicit form to $\Psi$ according to our purposes.

\section{Missing links}
\label{SecMiss}

Recall that our model assumes that no death occurs, which, as we have seen, is consistent with the fact that the work is related to perennial plant cultivars with asexual multiplication. However, individuals are obviously missing in the population -- because they represent intermediate individuals never recorded as a cultivar and never distributed by the breeder, because the cultivar disappeared from rose gardens deliberately or accidentally, or because it was not sampled in the study. In this section, our objective is to look for some \textit{missing links}. Since we do not know exactly how many individuals are missing, our strategy is to launch a \textit{greedy algorithm} that explores the population and tries to detect an excess of information that might improve substantially the genealogy. The combinatorial complexity leads us to focus on some particular areas for the algorithm. More precisely, it seems that the isolated individuals are suitable starting points, for which we recall that $\cI = \{ e \in \cP\, \vert\, c^{\, *}(e) = \varnothing \}$ is the set of individuals having no parents in the most likely genealogy. For all $e \in \cI$, let $\cR(e) \subset \cP$ be the individuals in the population chronologically candidates to the genealogy of $e$ and able to produce a gamete compatible with $e$. In addition, for each $p \in \cR(e)$, consider
\begin{equation}
\label{MissIndiv}
i^{*}(e,p) = \underset{i}{\arg \max} ~ \delta(e, p, i)
\end{equation}
as it is defined in \eqref{ProbaIndiv}, where $i$ has the structure of an individual of the population (with a ploidy, a date of birth and a set of alleles for each signal). Namely, $i^{*}(e,p)$ is a virtual individual having a genotype which maximizes the probability of the event $\{ (p, i) \mapsto e \}$, it can be seen as the ``perfect partner" of $p$ to produce $e$. Given $i = i^{*}(e,p)$, we now have to decide whether $i$ significantly improves the genealogy. Let us carry on with the criterion introduced in \eqref{LogLikPenGen}, where the enhanced population is $\cP_{i} = \cP \cup \{ i \}$. To match with our study, the penalization $\Psi(i)$ must favor individuals $i$ providing the maximum number of interactions with $\cP$. As we have seen in the last section, few interactions leave the likelihood almost unchanged whereas too many interactions tend to depreciate it, this was our motivation to look for a trade-off. We also want to give priority to any individual $i$ reducing the number of \textit{connected components} in the genealogy -- that is, the number of subgraphs in which all nodes are connected. Indeed, in view of our fundamental hypothesis that, except for ancestors, all parents should be present in an ideal population, we know that if we were able to access to the whole population, it would lead to a graph with few connected components (less than the number of ancestors, in any case). In this context, it seems natural to favor the reduction of the number of connected components, in order to get closer of this true (but inaccessible) genealogy. Define $\cT$ and $\cT_{i}$ as the maximum likelihood trees on $\cP$ and $\cP_{i}$, respectively, and suppose that $i$ is contained in $\cT_{i}$. Combining these requirements, we can write the penalization in the form
\begin{equation}
\label{LogLikPen}
\Psi(i) = \lambda_{i}\, \frac{r(i)}{n} - \mu_{i}\, \Delta C(i)
\end{equation}
where $r(i)$ is the number of individuals of $\cP$ potentially interacting with $i$, $\Delta C(i)$ is the difference between the number of connected components in $\cT$ and $\cT_{i}$, $\lambda_{i} \geq 0$ and $\mu_{i} \geq 0$ are regularization parameters. Our decision rule consists in keeping an individual $i$ which satisfies $\ell^{\, *}(\cT_{i}) > \ell(\cT)$. We can formalize $r(i)$ like
\begin{equation*}
r(i) = \sum_{p\, \in\, \cP} \eta(i,p)
\end{equation*}
where $\eta(i,p) = 1$ if one can find $a \in \cP$ such that $\delta(i, a, p) > 0$, $\delta(a, i, p) > 0$ or $\delta(p, a, i) > 0$, that is, if there is a nonzero probability for at least a link involving $p$ and $i$, and $\eta(i,p) = 0$ otherwise. Note that $a$ may be an offspring of $i$ as well as a parent or a partner of $i$ to be considered as an interaction involving $i$. To adapt our criterion, we can choose
\begin{equation}
\label{LogLikPenEstL}
\lambda_{i} = \frac{n}{2}\, \vert \ell(\cT) - \ell(\cT_{i}) \vert
\end{equation}
since this guarantees that $\ell^{\, *}(\cT_{i}) = \ell(\cT)$ when the new individual does not bring any connection except the one for which it has been created, not gathering connected components ($r(i) = 2$ and $\Delta C(i) = 0$), and thus when $i$ should be rejected. A similar strategy enables us to fix $\mu_{i}$, for $r(i)=2$ must at least coincide with $\Delta C(i) = -1$ to make an interesting link. This is the case when $i$ has been created to fulfill the event $\{ (p, i) \mapsto e \}$, and when $p$ and $e$ belong to different connected components. Of course that situation must be favored, and to simplify one can choose
\begin{equation}
\label{LogLikPenEstM}
\mu_{i} = \lambda_{i} + 1
\end{equation}
which amounts to say that $\ell^{\, *}(\cT_{i}) > \ell(\cT)$ whenever $\Delta C(i) < 0$. To enhance the population, we suggest the following algorithm.
\begin{enumerate}
\setcounter{enumi}{-1}
\item \label{Step0} Fix $n_{v} > 0$, the maximum number of virtual individuals allowed to be inserted in the population.
\item \label{Step1} Build $\cR(e)$ for all $e \in \cI$.
\item \label{Step2} For all $p \in \cR(e)$, compute the maximum likelihood partner $i$ such that $\{ (p, i) \mapsto e \}$ is achieved.
\item \label{Step3} Among these candidates, add in $\cP$ the individual maximizing $\ell^{\, *}(\cT_{i})$ provided
$$
\max_{i} ~ \ell^{\, *}(\cT_{i}) > \ell(\cT).
$$
Set $t_{e}-1$ as birth date of the new individual, where $t_{e}$ is the one of $e$.
\item \label{Step4} Recalculate the most likely tree $\cT$ and the set $\cI$ according to the new population.
\item Repeat steps \eqref{Step1}--\eqref{Step4} as long as the criterion increases and $\card(\cP) < n+n_{v}$.
\end{enumerate}

\smallskip

Before going further, let us focus on the complexity of this algorithm (and on some possible improvements). In the present state, it is fully exploratory and starts from arbitrary points. In terms of complexity, it is possible to evaluate that step \eqref{Step4} has a number of crosses in the range of $O(n (n-1) (n-2))$ to be tested. Generally $\card(\cI)$ is small, thus, even if it entirely depends on the population, let us suppose that it is bounded by $n_{i} \ll n$. The construction of $\cR(e)$ requires $O(n-1)$ crosses to be tested for a given $e$. On the whole, we can roughly estimate that, considering a crossbreeding as the unit of measurement, $O(n_{v}\, n_{i}\, n (n-1)^2(n-2))$ operations are needed. In practice, much less operations are actually done since the symmetry and the chronological and genetical constraints cut a lot of paths. To reach a lower complexity, it should be relevant to look at less exploratory methods, in order to deal with the increasing number of individuals. In addition, the maximum of likelihood in step \eqref{Step2} is the natural solution, but it can also have unwelcome effects. In particular, this algorithm can not generate any triploid. This follows from the fact that, whenever a triploid produces a gamete, there exists a diploid or a tetraploid that produces the same gamete with a probability two times bigger. In the same vein, the virtual tetraploids can either be homozygous or heterozygous with only two distinct alleles. As a consequence, since the individual is specifically created to fulfill a particular crossbreeding, the situations where the missing link is a parent of more than one offspring in the population can not be recovered, except if the offsprings are genetically similar. This could be improved by testing not only the candidates, but also the mixes between them. For example, if $\{ a,a,b,b \}$ is added to explain the presence of a diploid $\{ a,b \}$, and if $\{ c,c,d,d \}$ is added to explain the presence of another diploid $\{ c,d \}$, then it could be interesting to add $\{ a,b,c,d \}$ to explain both of them, instead. To conclude, we would like to highlight a last enhancement. Setting $t_{e}-1$ as date of birth of the new individual is an arbitrary choice because, focusing on the offspring, we do not have any more information about the other interactions within the new genealogy. Each birth date between some initial time $t_0$ and $t_{e}-1$ should be tested as well. All these improvements are hardly conceivable due to the computational complexity, except for small populations ($n \approx 50$, as in our simulations). Hence, as we can see, there are still numerous open questions to explore on the fundamental issue of the missing links.

\section{An empirical study}
\label{SecEmp}

The numerical processings were carried out through the \texttt{R} programming language and its software environment. In particular, we used the package \textit{igraph}\footnote{\texttt{https://cran.r-project.org/web/packages/igraph/igraph.pdf}} to display the graphs. In all figures of this section, the geometric shapes that we use are circles to represent diploids, triangles for triploids and squares for tetraploids, gray individuals are real whereas white individuals are virtual. Similarly, we use solid lines for true links as well as dotted lines for the wrong links given by the model (unless noted otherwise). The computations are conducted \textit{via} the uniform probabilities $\pi_{21} = \pi_{12} = \frac{1}{2}$ and $\pi_{31} = \pi_{22} = \pi_{13} = \pi_{211} = \pi_{121} = \pi_{112} = \frac{1}{3}$. The estimation of the mean number of offsprings is given by \eqref{EstEspSelect} and the \textit{outlier threshold} is chosen to the standard $q_3 + 1.5\, (q_3-q_1)$ with $q_1$ and $q_3$ the first and third quartiles of a subset of observations. It is computed using a moving window on the values in chronological order and then extrapolated by a linear regression, to take into account the time-invariance in the reproduction law and, thus, the fact that the older an individual is, the more offsprings he is likely to have.

\subsection{On a simulated population}
\label{SecEmpSim}

\begin{figure}[h!]
\includegraphics[width=16.5cm]{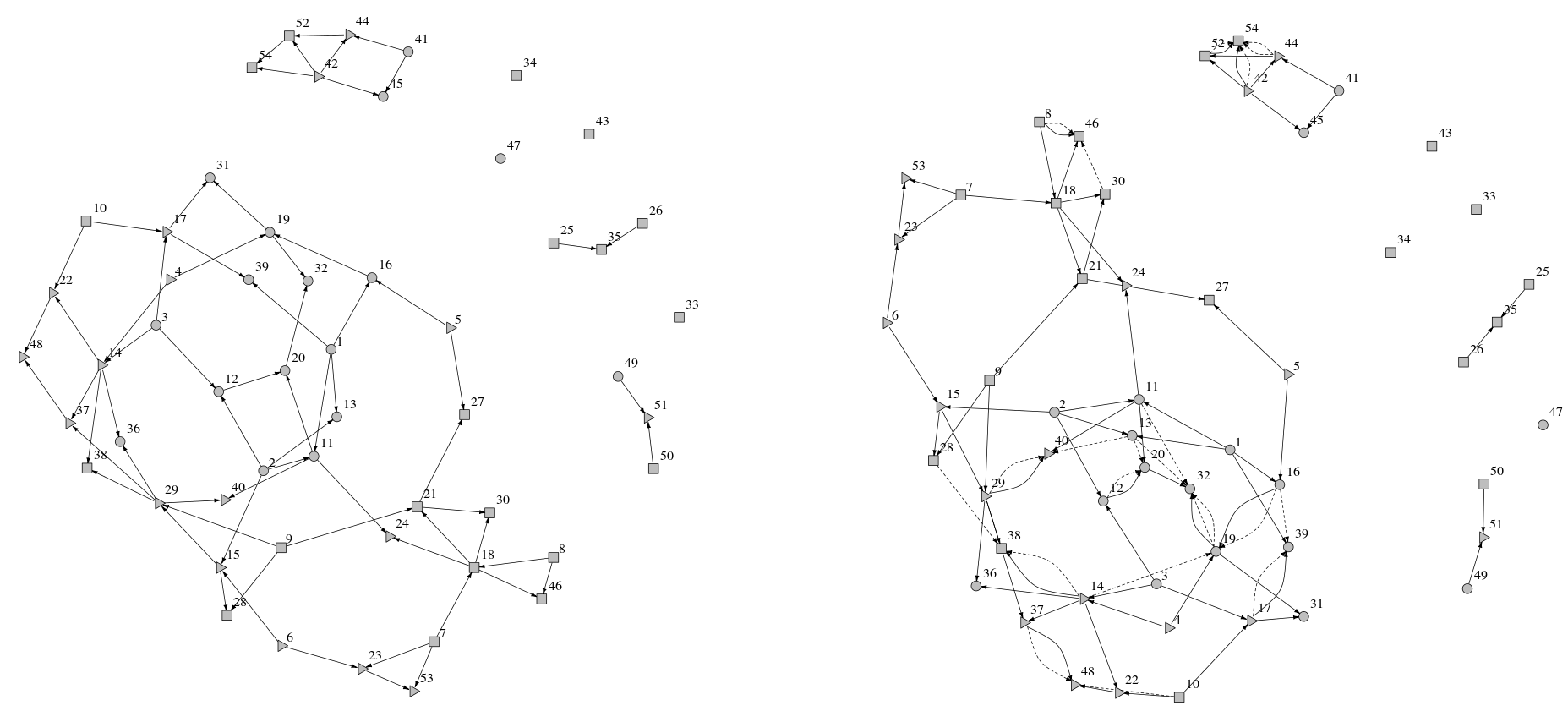}
\caption{True genealogy $\cT^{\, 0}$ of the simulated population, on the left. Superposition of all genealogies of the simulated population found by the model, on the right.}
\label{FigTrueFullGenSim}
\end{figure}

Consider the simulated population $\cP$ whose detailed description is provided in the Appendix. To sum up, there are $n=54$ individuals among which 17 diploids, 17 triploids and 20 tetraploids have interacted throughout 8 generations. The simulation relies on $m=4$ genes, dates of birth are known (\textit{via} the generations) as are ploidies and observed genotypes. The goal is to apply our model on this population and to put the results into perspective, compared with the true genealogy $\cT^{\, 0}$ which is represented on the left of Figure \ref{FigTrueFullGenSim}.

\begin{figure}[h!]
\includegraphics[width=16.5cm]{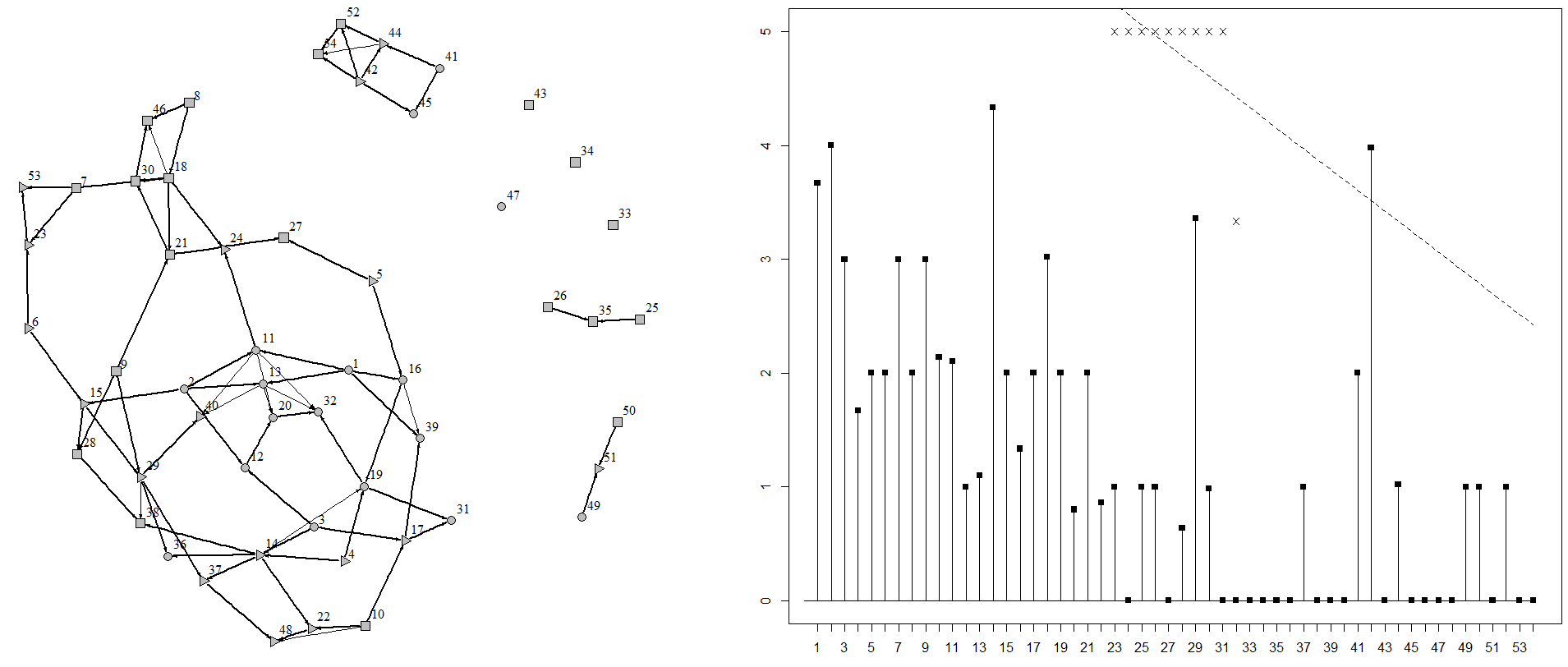}
\caption{Genealogical graph of the simulated population, on the left. The thickness of the links is proportional to their weights in the model. Mean number of offsprings for each individual, on the right. The abscissa displays the individuals $i \in \cP$ in chronological order and the ordinate represents the estimated expectation of $N(i)$. The dotted line is the outlier threshold extrapolated from the crosses (the moving window goes through 22 observations). There is 1 probably favored individual.}
\label{FigGraphGenSim}
\end{figure}

\subsubsection{Family trees and most likely genealogy} 
\begin{figure}[h!]
\includegraphics[width=16.5cm]{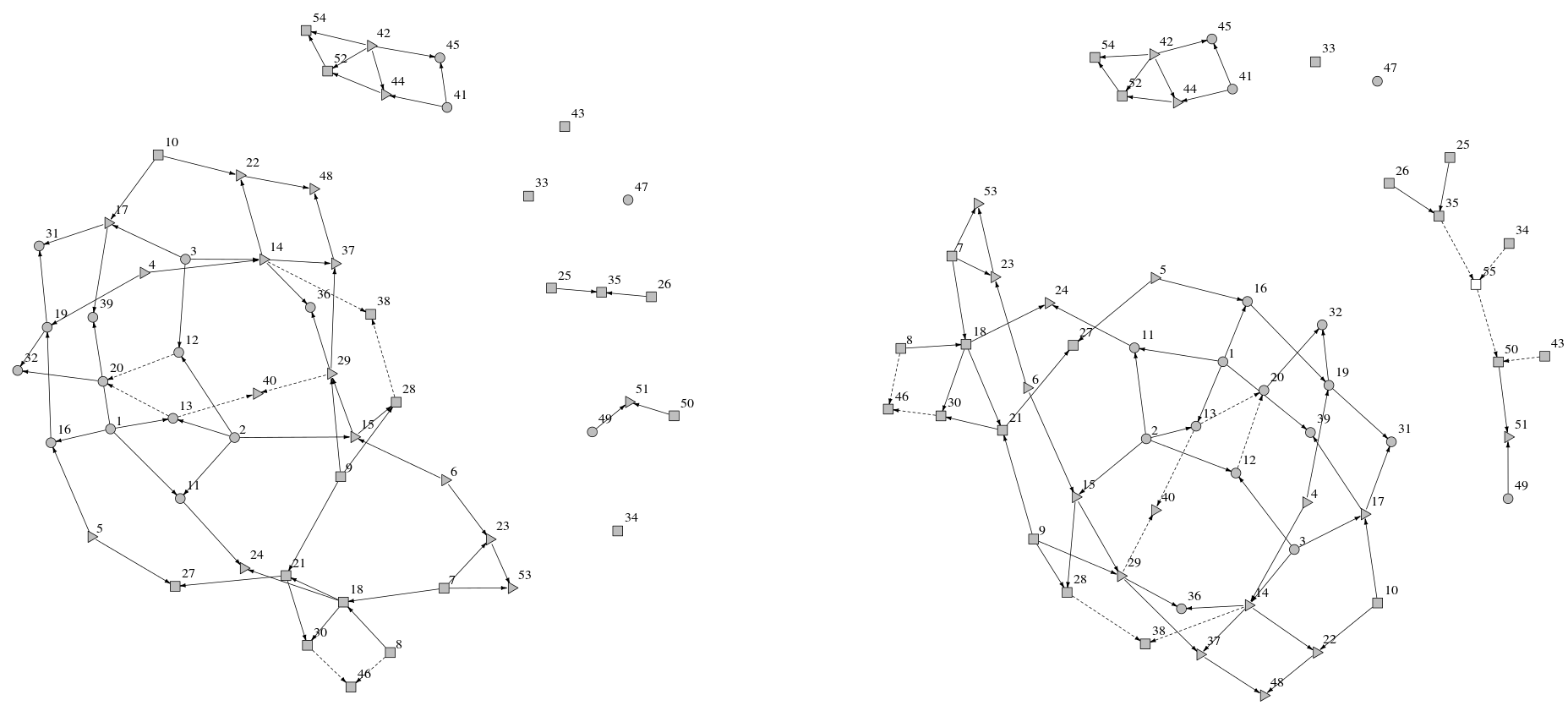}
\caption{Genealogy $\cT^{\, *}$ maximizing the log-likelihood of the simulated population found by the model, on the left. There are 8 connected components. Genealogy $\cT_1$ maximizing the log-likelihood of the simulated population enhanced with one individual (55) found by the model, on the right. There are 5 connected components.}
\label{FigMaxNouv1GenSim}
\end{figure}

All genealogies found by the model have been superposed on the right of Figure \ref{FigTrueFullGenSim}, that is, the full content of $G(e)$ given in \eqref{ProbaIndivFull} for each $e \in \cP$. Similarly, we have also added in Figure \ref{FigGraphGenSim} the genealogical graph of the population as it is defined in \eqref{GenGraph}, highlighting the pairwise potential relationships. We can first verify that the ancestors (individuals from 1 to 10) are only parents. On the one hand, we observe that the true genealogy is included in the graph, illustrating thereby the effectiveness of the exploratory algorithm. One can also notice, on the other hand, that some wrong links have been detected. We should however indicate that a wrong link is not an \textit{impossible} link, for the reader can check that dotted arcs correspond to compatible crosses. Consider as an example the link $\{ (14, 28) \mapsto  38 \}$ appearing in Figure \ref{FigTrueFullGenSim} but absent from the true genealogy. We have $x(14) = 3$, so  $\wh{g}_2(14) = \{ 160, 170, 180 \} = g_2(14)$. Similarly, with $x(28) = 4$ and $x(38) = 4$, $\wh{g}_2(28) = \{ 210, 290 \}$ can correspond to $g_2(28) = \{ 210, 290, 290, 290 \}$ and $\wh{g}_2(38) = \{ 160, 170, 290 \}$ to $g_2(38) = \{ 160, 170, 290, 290 \}$. Through pattern (P$_{\ref{Sch34}}$), a genealogical link is possible on the signal 2 and we easily check that the same conclusion holds on each signal. This is an illustration of the fact that, from a practical point of view -- namely, with an \textit{unknown} true genealogy -- it is preferable to produce a set of possible genealogies instead of a single one. Afterwards, the accumulation of genes enables pruning of the trees, step by step, to reinforce the remaining branches. To support this argument, Figure \ref{FigMaxNouv1GenSim} shows on its left the family tree $\cT^{\, *}$ maximizing the log-likelihood \eqref{LogLikGen} in which we observe that the true genealogy was \textit{not} the most likely one, retrospectively. Let us have a look at the differences. The first one is the selection of $\{ (14, 28) \mapsto  38 \}$ instead of $\{ (14, 29) \mapsto  38 \}$. Knowing that 28 and 29 both have parents $(9,15)$, we easily understand their genetic likeness. The second one is interpreted in the same way since $\{ (8, 30) \mapsto  46 \}$ stands in for $\{ (8, 18) \mapsto  46 \}$, and since 18 is a parent of 30. For the last two ones, 13 takes the place of 11 in the true connections $\{ (11, 12) \mapsto  20 \}$ and $\{ (11, 29) \mapsto  40 \}$, 11 and 13 having the same parents. To be precise, in the latter example each link leads to the same probability and the maximum of likelihood is not unique (in which case the algorithm chooses one solution at random). On this dataset, we get
\begin{equation*}
\ell(\cT^{\, *}) \approx -3.052 ~ > ~ -7.616 \approx \ell(\cT^{\, 0}).
\end{equation*}
Even so, wrong links maximizing the log-likelihood are usually relevant. In this example, the wrong parents detected are in fact close relatives of true parents. To sum up the results of this simulation, amongst the 45 potential triplets that form the full genealogies, 34 are true and 11 are wrong, but all true links are correctly retrieved. In the maximum likelihood genealogy, one can find from 30 to 32 true links and from 2 to 4 wrong links. The two links that can either be true or wrong have equal probabilities, as it has just been detailed. Even if it is of lesser interest on a simulation, Figure \ref{FigGraphGenSim} also contains the estimated expectations of the number of offsprings in the population, on the basis of all genealogies with no threshold $(\pi_{\min} = 0)$. The individual 42 appears as favored and, indeed, one can check that it has 4 offsprings in the true genealogy whereas it belongs to generation 5. In terms of mean error between the estimated number of offsprings $\wh{\dE}[N(i)]$ and the number of offsprings $n^{*}(i)$ in the maximum likelihood genealogy,
\begin{equation*}
\frac{1}{n} \sum_{i\, \in\, \cP} \big\vert \wh{\dE}[N(i)] - n^{*}(i) \big\vert \approx 8.52 \times 10^{-2} \hsp \text{and} \hsp \frac{1}{n} \sum_{i\, \in\, \cP} \big( \wh{\dE}[N(i)] - n^{*}(i) \big)^2 \approx 5.19 \times 10^{-2}.
\end{equation*}

\subsubsection{Missing links} 
\begin{figure}[h!]
\includegraphics[width=16.5cm]{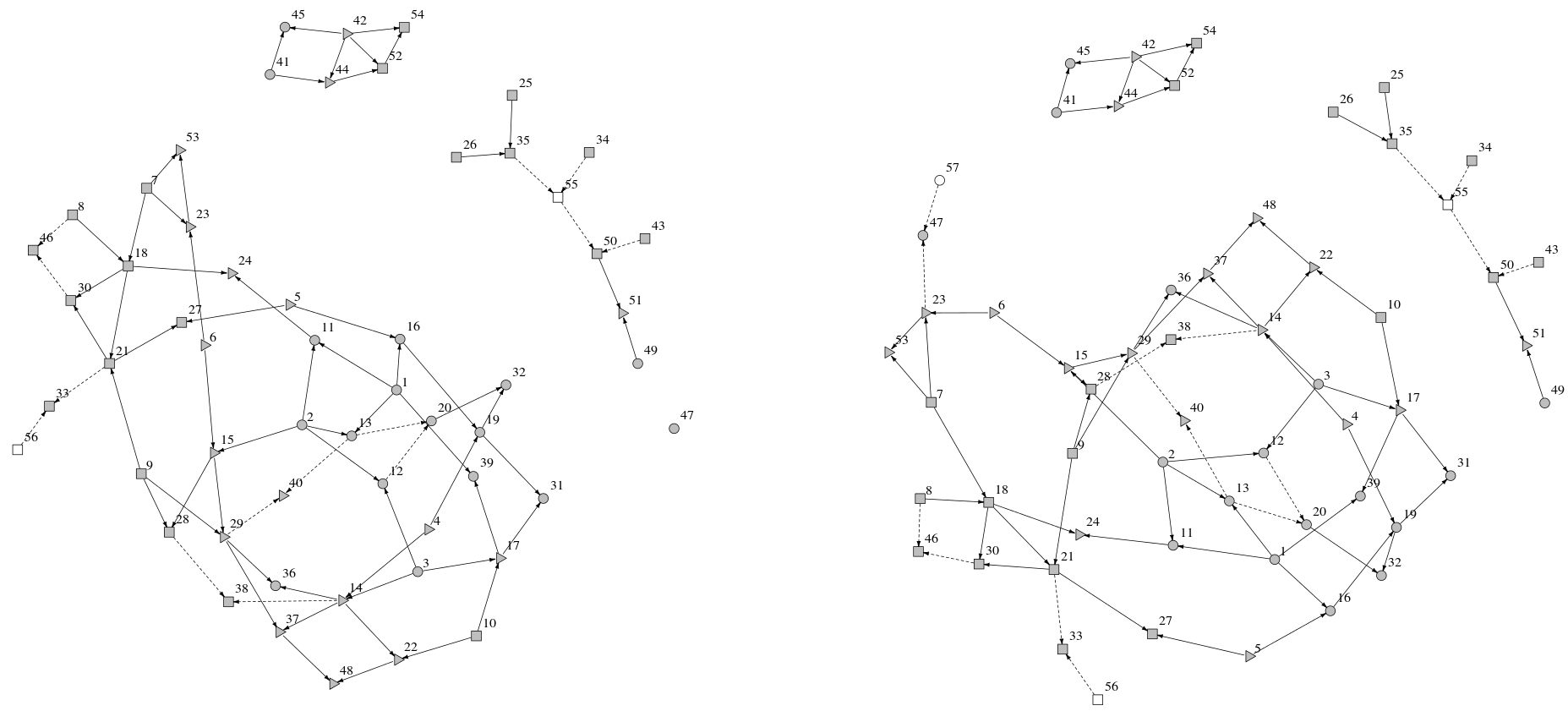}
\caption{Genealogy $\cT_2$ maximizing the log-likelihood of the simulated population enhanced with two individuals (55 and 56) found by the model, on the left. There are 4 connected components. Genealogy $\cT_3$ maximizing the log-likelihood of the simulated population enhanced with three individuals (55, 56 and 57) found by the model, on the right. There are 3 connected components.}
\label{FigNouv2Nouv3GenSim}
\end{figure}

We now look for missing links, following the algorithm described at the end of Section \ref{SecMiss} with $n_{v} = 3$. Compared with the most likely tree $\cT^{\, *}$ on the population $\cP$, the largest increase of our penalized criterion $\ell^{\, *}$ given by \eqref{LogLikPenGen} is reached by adding the tetraploid  $g_1(55) = \{ 200, 200, 200, 200 \}$, $g_2(55) = \{ 270, 270, 270, 270 \}$, $g_3(55) = \{ 370, 370, 370, 370 \}$ and $g_4(55) = \{ 410, 410, 520, 520 \}$,  respectively for the 4 genes, as a member of generation 5. We obtain the genealogy on the right of Figure \ref{FigMaxNouv1GenSim}. From 8 connected components in $\cT^{\, *}$, only 5 remain in the maximum likelihood tree $\cT_1$ on the population enhanced with the individual 55 having this precise genotype. Thus its role as a missing link is clearly highlighted and that explains the reason why it has been privileged, even if $\ell(\cT_1) \approx -3.106$ has decreased compared to $\ell(\cT^{\, *}) \approx -3.052$. A second loop of the algorithm generates the tetraploid having $g_1(56) = \{ 10, 10, 200, 200 \}$, $g_2(56) = \{ 130, 130, 380, 380\}$, $g_3(56) = \{ 210, 210, 370, 370 \}$ and $g_4(56) = \{ 430, 520, 520, 520 \}$ on its 4 genes, in generation 3. Only 4 connected components remain, but the log-likelihood is now $\ell(\cT_2) \approx -3.482$. The last loop of the algorithm gives a diploid $g_1(57) = \{ 90, 90 \}$, $g_2(57) = \{ 220, 220 \}$, $g_3(57) = \{ 310, 310 \}$ and $g_4(57) = \{ 510, 510 \}$ in generation 5. Only 3 connected components remain while, for this last addition, the log-likelihood is unchanged. Figure \ref{FigNouv2Nouv3GenSim} depicts $\cT_2$ and $\cT_3$, respectively on the left and on the right. This simulated example seems to clearly illustrate the operation of the exploratory algorithm, focusing on connected components to build missing links, retrospectively. To support the remarks of Section \ref{SecMiss} about the algorithm, suppose now that the diploid 49 is removed from the dataset. Then, amongst all virtual candidates, a new diploid -- say $49^{*}$ -- with genotype $\{ 240, 240 \}$, $\{ 320, 320 \}$, $\{ 410, 410 \}$ and $\{ 410, 410 \}$ appears in generation 6. One can check that this does not correspond to the real 49, but this new genotype allows the cross $\{ (49^{*},50) \mapsto 51 \}$ with a bigger probability than what actually occurred (precisely, $\frac{1}{2} \times 1 \times \frac{1}{6} \times \frac{1}{12} < \frac{1}{2} \times 1 \times \frac{1}{6} \times \frac{1}{6}$). From this point of view, the algorithm is consistent since there is no way we can retrieve the true allele 510 instead, not spread elsewhere. However, if the diploid 1 is removed from the dataset, then, because it is involved in numerous relationships and because it is heterozygous in most cases, a unique individual playing the same roles is not recovered. For example, on signal 1 and 4, alleles 20 and 310 are needed for $\{ (1,2) \mapsto 13 \}$ whereas 10 and 320 are needed for $\{ (1,2) \mapsto 11 \}$. The algorithm suggests an individual $\{20, 20 \}$ and $\{ 310, 310 \}$ and another one $\{ 10, 10 \}$ and $\{ 320, 320 \}$ on these signals, because they maximize the likelihood of the crossbreedings with 2 to produce 11 and 13. In the end, all genetic information is retrieved but, to be improved, the process should also mix the candidates beforehand, considering $\{ 10, 20\}$ and $\{ 310, 320 \}$ in this case, as we have mentioned it in the enhancements.

\subsection{On a rose bushes population}
\label{SecEmpRos}

\begin{figure}[h!]
\includegraphics[width=13cm]{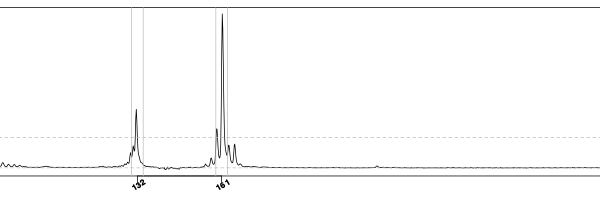}
\caption{Example of signal for a particular microsatellite marker. The individual $i$ is tetraploid and two peaks have been detected. Here $\wh{g}(i)$ is $\{ 132, 161 \}$ and $g(i)$ is $\{ 132, 132, 132, 161 \}$ with probability $\pi_{31}$, $\{ 132, 132, 161, 161 \}$ with probability $\pi_{22}$ and $\{ 132, 161, 161, 161 \}$ with probability $\pi_{13}$. To simplify, scales are deliberately removed.}
\label{FigExSig}
\end{figure}

To conclude the study, we are now going to launch our model on a subpopulation of rose bushes collected on the basis of $m=4$ genes. We start by giving some explanations about the experimental gathering of the data. Among molecular markers, microsatellite markers are still a reference for pedigree reconstruction because they are highly multiallelic codominant markers \cite{Jones2003}. After Polymerase Chain Reaction (PCR), amplified fragments are generally separated by capillary electrophoresis. According to their size, amplified fragments are detected at a given time of the electrophoresis and are depicted as a peak in the electrophoregram, whose area varies according to the intensity of the signal. Thus, a statistical treatment of the four signals of the individual $i$ gives the observed genotypes $\wh{g}(i)$. To deal with allelic multiplicity, theoretical ratios between peak intensities could be used to determine the relative number of copies of each allele in polyploids \cite{Esselink2004}. Unfortunately this strategy is very difficult to apply, especially because signal intensity is also dependent on amplification competition between alleles during PCR. Therefore, in most cases electrophoregrams are generally interpreted as presence or absence of alleles \cite{Dufresne2014}. This is also our approach in this article but considering all possibilities of multiplicity, for which we have seen in the previous sections how our model enables building and probabilizing of $g(i)$ from $\wh{g}(i)$. An example of signal is shown in Figure \ref{FigExSig}. In addition we must not forget that a calibration of the equipment is needed, for practical purposes. In concrete terms, the abscissa of the signals is made of decimal values, which is clearly incompatible with what it is supposed to highlight, namely some \textit{base pairs}. Hence we take rounded values, and an offset of $\pm 1$ for each allele has to be considered. This is the reason why we decided to switch to criterion \eqref{EqGenArr} in the real data analysis.

\subsubsection{Family trees and most likely genealogy} 
\begin{figure}[h!]
\includegraphics[width=10cm]{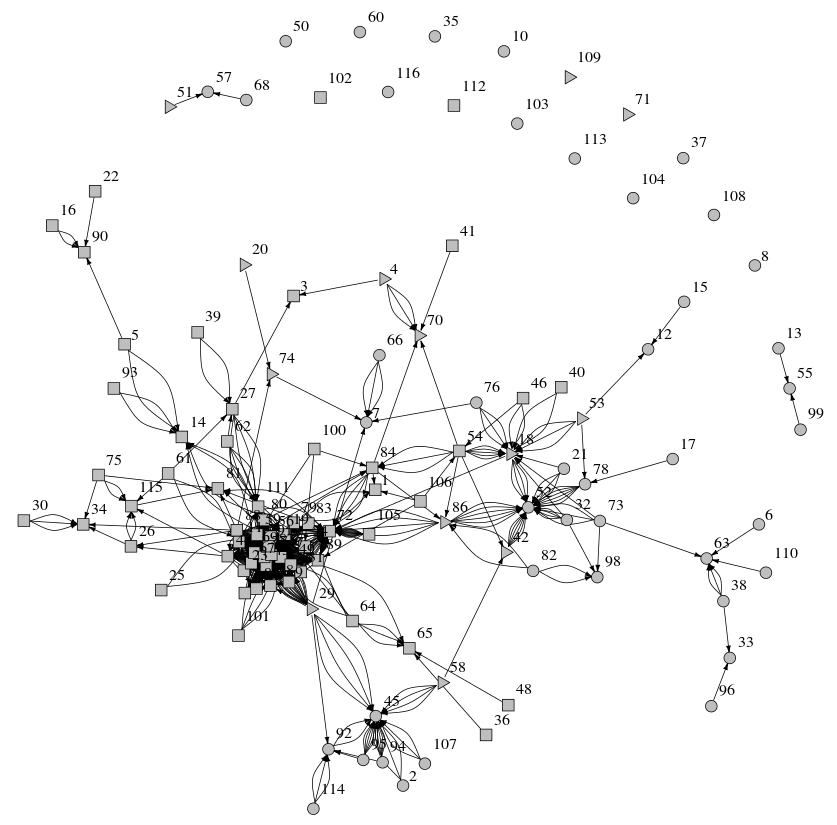}
\caption{Superposition of all genealogies of the rose bushes subpopulation found by the model.}
\label{FigFullGenReal}
\end{figure}

\begin{figure}[h!]
\includegraphics[width=10cm]{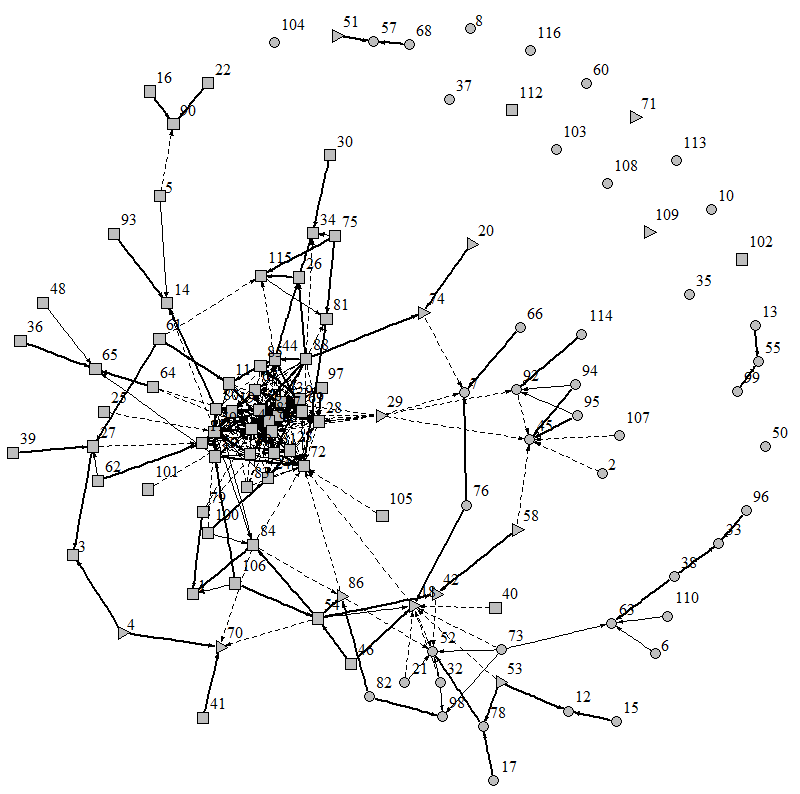}
\caption{Genealogical graph of the rose bushes subpopulation. The thickness of the links is proportional to their weights in the model. The dotted lines correspond to potential links set to zero by the threshold probability.}
\label{FigGraphGenReal}
\end{figure}

Now we put aside $n=116$ rose bushes, selected for the knowledge of their ploidy and for the clarity of their signals, and we look for potential genealogical links among them using the same allelic probabilities as in the simulation study. All genealogies are superposed on Figure \ref{FigFullGenReal} together with the genealogical graph on Figure \ref{FigGraphGenReal} for the threshold probability $\pi_{\min} = 0.2$, a choice that will be justified in the sequel. Even if the graphical representation seems unexploitable, it illustrates the fact that many solutions are conceivable. More than one genealogy maximizes the likelihood, for some links have the same probability. An example of most likely genealogy is given on the left of Figure \ref{FigMaxMissGenReal}, it contains 35 connected components. Within the largest one, a chain of 5 generations is obtained ($9 \rightarrow 56 \rightarrow 67 \rightarrow 59 \rightarrow 47$).

\subsubsection{Missing links} 
\begin{figure}[h!]
\includegraphics[width=16.5cm]{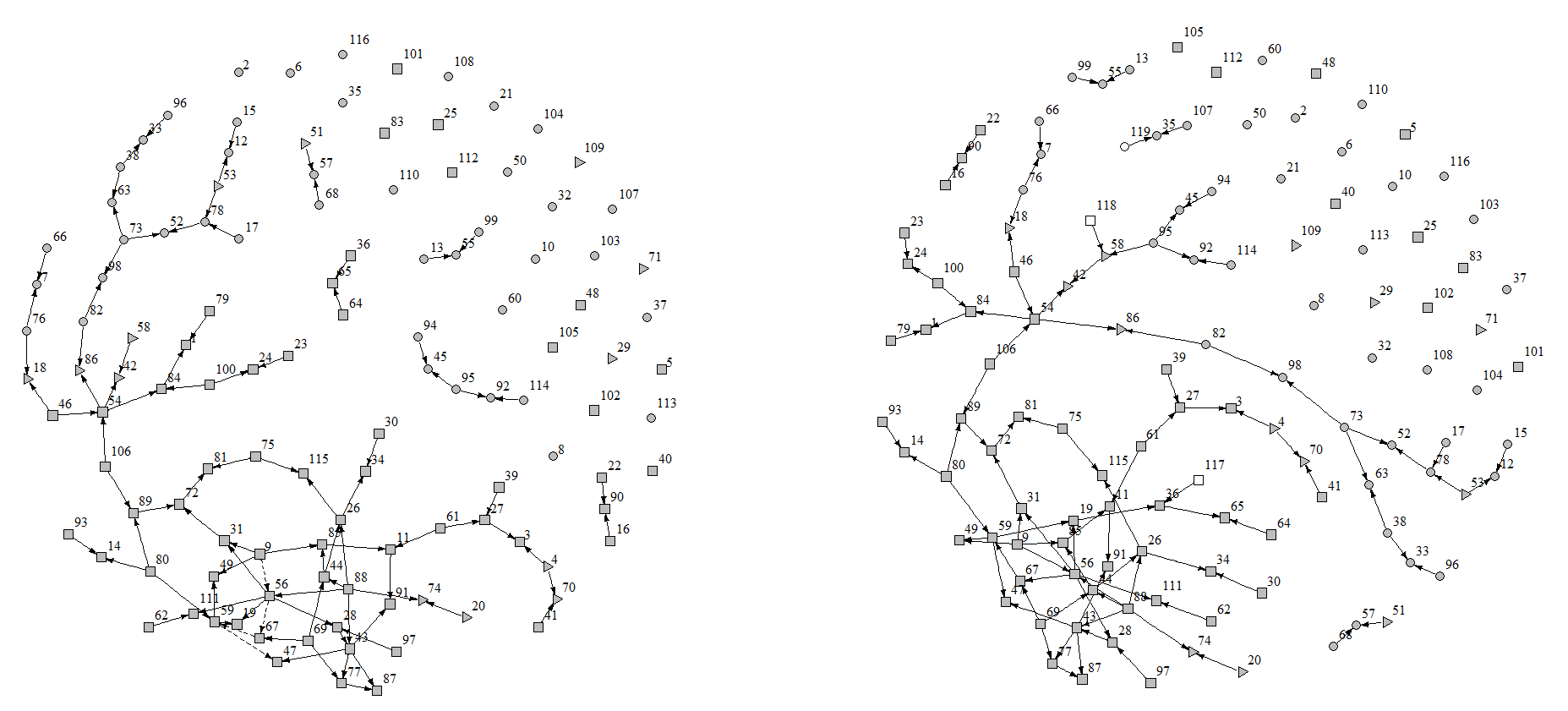}
\caption{Genealogy $\cT^{\, *}$ maximizing the log-likelihood of the rose bushes subpopulation found by the model (the dotted line highlights a chain of 5 generations), on the left. There are 35 connected components. Genealogy $\cT_3$ maximizing the log-likelihood of the rose bushes subpopulation enhanced with three individuals (117, 118 and 119) found by the model, on the right. There are 32 connected components.}
\label{FigMaxMissGenReal}
\end{figure}

On the right of Figure \ref{FigMaxMissGenReal}, one of the most likely genealogies is represented when $n_{v} = 3$ new individuals suggested by the algorithm of Section \ref{SecMiss} are added (117, 118 and 119). Again, their role as missing links and their usefulness to connect separated branches of the genealogy are clearly brought to light. Only 32 of them remain, due to the fact that each missing link connects two components. In particular, we can notice the important intercession of 118, plugging the two largest ones.

\subsubsection{Selected individuals} 
\begin{figure}[h!]
\includegraphics[width=16.5cm]{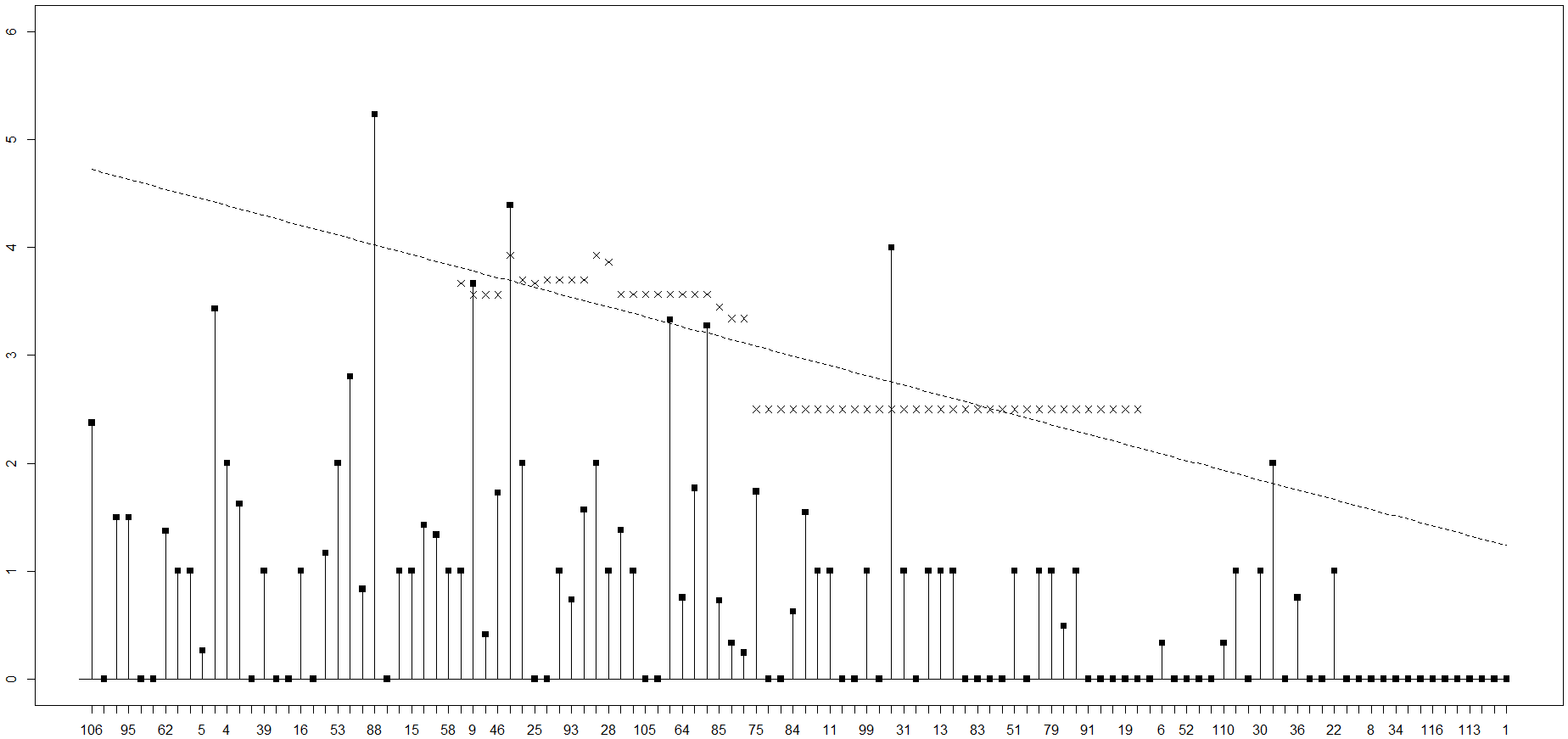}
\caption{Mean number of offsprings for each individual. The abscissa displays the individuals $i \in \cP$ in chronological order and the ordinate represents the estimated expectation of $N(i)$. The dotted line is the outlier threshold extrapolated from the crosses (the moving window goes through 30 observations). For readability reasons, the abscissa is not completely filled. There are 6 probably favored individuals.}
\label{FigEstEsp}
\end{figure}

\begin{figure}[h!]
\includegraphics[width=12cm]{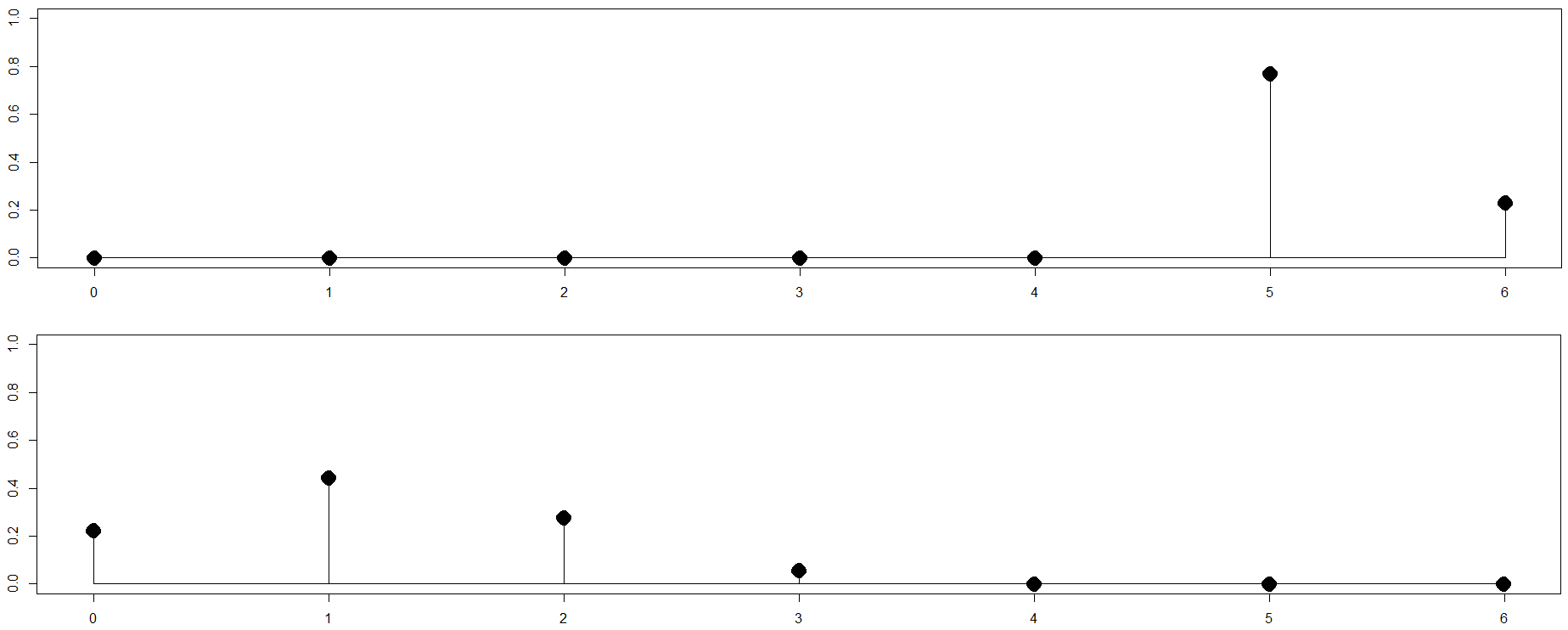}
\caption{Empirical distribution of the random variable $N(88)$, at the top. The abscissa represents the number $k$ of offsprings, the ordinate is the estimated probability associated with the event $\{ N(88) = k \}$. At the bottom, empirical distribution of the random variable $N(73)$.}
\label{FigEstProba}
\end{figure}

To look for selected individuals, the estimated probabilities \eqref{EstProbaSelect} and expectations \eqref{EstEspSelect} are computed for all $i \in \cP$ on the basis of a subset of genealogies made of links whose likelihood is greater than $\pi_{\min} = 0.2$. Indeed, since $\card(\dG(\cP)) > 10^{28}$ the computation with no threshold is infeasible. It appears that with this choice of threshold, $\card(\dG(\cP))$ is in the range of $10^{6}$ which is small enough to proceed to computations and large enough to trust the statistical estimations. Figure \ref{FigEstEsp} contains the empirical expectations of all individuals together with an outlier threshold, evaluated as it is explained in the beginning of this section. Each individual having a higher mean number of offsprings is considered as a potential target for the retrospective selection by breeders, there are 6 in this subpopulation. Amongst all individuals, $i=88$ has, on average, the largest number of offsprings in the population. Figure \ref{FigEstProba} shows the empirical distribution of $N(88)$. Concretely,
\begin{equation*}
\wh{\dP}(N(88) = 5) \approx 0.770, \hsp \wh{\dP}(N(88) = 6) \approx 0.230 \hsp \text{and} \hsp \wh{\dE}[N(88)] \approx 5.230.
\end{equation*}
The last empirical distribution represented is the one of $N(73)$, chosen to illustrate the fact that an individual may have offspring in some genealogies and no offspring in the others. Numerically,
\begin{equation*}
\wh{\dP}(N(73) = 0) \approx 0.222, \hsp \wh{\dP}(N(73) = 1) \approx 0.444, \hsp \wh{\dP}(N(73) = 2) \approx 0.278,
\end{equation*}
\begin{equation*}
\wh{\dP}(N(73) = 3) \approx 0.056 \hsp \text{and} \hsp \wh{\dE}[N(73)] \approx 1.167.
\end{equation*}
In terms of mean error between the estimated number of offsprings $\wh{\dE}[N(i)]$ and the number of offsprings $n^{*}(i)$ in the maximum likelihood genealogy,
\begin{equation*}
\frac{1}{n} \sum_{i\, \in\, \cP} \big\vert \wh{\dE}[N(i)] - n^{*}(i) \big\vert \approx 1.21 \times 10^{-1} \hsp \text{and} \hsp \frac{1}{n} \sum_{i\, \in\, \cP} \big( \wh{\dE}[N(i)] - n^{*}(i) \big)^2 \approx 7.30 \times 10^{-2}.
\end{equation*}

\section{Conclusion}
\label{SecConclu}

To conclude, we would like to draw the attention of the reader to some weaknesses of the model, essentially relying on the allelic multiplicity. Indeed, our choice of considering each potential multiplicity weighted by a probability, instead of selecting a particular one, may lead to contradictions in the genealogy. Suppose for simplification that the most likely genealogy contains the links $\{ (p_1, p_2) \mapsto q_1 \}$ and $\{ (q_1, q_2) \mapsto e \}$ where $p_1$ is a tetraploid such that $g(p_1) = \{ a,a,a,a \}$, and $p_2$ is a diploid such that $g(p_2) = \{ b,b \}$. Both of them are homozygous, so there is no allelic uncertainty derived from their observed genotypes, but $\wh{g}(q_1) = \{ a,b \}$ for the triploid $q_1$ can only match with $\{ (p_1, p_2) \mapsto q_1 \}$ in case of $g(q_1) = \{ a,a,b \}$. Suppose now that $q_2$ and $e$ are tetraploids, having $g(q_2) = \{ c,c,c,c \}$ and $\wh{g}(e) = \{ b,c \}$, respectively. Then, the link $\{ (q_1, q_2) \mapsto e \}$ has a nonzero probability only for $g(q_1) = \{ a,b,b \}$. In other words, the most likely genealogy treats $q_1$ as a link between $(p_1,p_2)$ and $e$, but at the cost of incompatible allelic combinations. This is a trail for future improvements of our model, in particular it seems worth considering an algorithm to detect contradictions and to eliminate such trees from the set of genealogies. Another weakness is the estimation of $\pi_{21}, \pi_{12}, \pi_{31}, \hdots$, namely the probabilities of allelic multiplicity. As we have seen in Section \ref{SecEmpRos}, we lack information to properly evaluate them. An ambitious track could be the generalization of \cite{Chaumont2017}, in which the authors establish the well-known \textit{Hardy-Weinberg equilibrium} to deal with heterozygoty in a diploid population. A challenging study will be to characterize this equilibrium in our polyploid population -- if it exists -- and to determine its degrees of freedom. This additional information will enable us to refine the probabilities of multiplicity, considering that the population has reached its equilibrium. The crossbreeding patterns also have to be enhanced with double reductions and preferential matches, both of them easily treated on a theoretical point of view (dealing with double reductions as rare events of probability $0 < \epsilon \ll 1$ and preferential matches as a lack of uniformity in the gamete production, when computing the probability of the crossbreeding), but difficult to estimate. We have widely discussed the algorithm for missing links and its status of working base which calls for numerous enhancements. Finally, it is important to insist upon the fact that this work is mainly theoretical and that the application of our model on a real population of rose bushes is only relevant in order to show that coherent and interpretable results are obtained. Nevertheless, we cannot draw any conclusion from an empirical study relying on $m=4$ genes. In-depth experiments will be conducted on more genes, and the comparison of any interesting result with available historical sources will constitute strong arguments to understand the breeders strategies over the past centuries, and also to try to complete the datasets with some lost or missing information.

\smallskip

\noindent \textbf{Acknowledgements}. This research was conducted in the framework of the regional programme ``Objectif V\'eg\'etal, Research, Education and Innovation in Pays de la Loire", supported by the French R\'egion Pays de la Loire, Angers Loire M\'etropole and the European Regional Development Fund, in the framework of the PedRo project. Empirical data were obtained thanks to the support of the R\'egion Pays de la Loire in the framework of the FLORHIGE project, by the National Institute of  Agricultural Research (INRA) and the National Natural Science Foundation of  China in the framework of  the SIFLOR project, and by the French  Ministry of Higher Education and  Research. The authors also thank the University of Bretagne Loire, Angers Loire M\'etropole and the University of Angers for their financial support, and the two anonymous reviewers for the numerous constructive comments and suggestions that  helped to improve substantially the paper.
\section*{Appendix}
\label{SecApp}

This Appendix is devoted to the precise description of the simulated population appearing in Section \ref{SecEmpSim}. All useful information are given in Tables \ref{TabSimPop12}, \ref{TabSimPop345} and \ref{TabSimPop678}, displaying the composition of the successive generations. For each individual, the columns indicate an identifier $i$, the ploidy $x(i)$, the observed genotypes $\wh{g}(i)$ on the four signals, the couple of parents and the reproduction pattern.

\begin{table}[h!]
\tiny
\begin{tabular}{|c|c|>{\centering}m{2.3cm}|>{\centering}m{2.3cm}|>{\centering}m{2.3cm}|>{\centering}m{2.3cm}|>{\centering}m{0.9cm}|c|}
\hline
\multicolumn{8}{|c|}{Generation 1} \\
\hline
$i$ & $x(i)$ & $\wh{g}_1(i)$ & $\wh{g}_2(i)$ & $\wh{g}_3(i)$ & $\wh{g}_4(i)$ & Par. & Pat. \\
\hline
1 & 2 & 10--20 & 110 & 210--310 & 310--320 & $\varnothing$ & -- \\
\hline
2 & 2 & 30--40 & 130--140 & 220--230 & 330 & $\varnothing$ & -- \\
\hline
3 & 2 & 50 & 150--160 & 240--250 & 340 & $\varnothing$ & -- \\
\hline
4 & 3 & 60 & 170--180--190 & 260--270 & 350--360--370 & $\varnothing$ & -- \\
\hline
5 & 3 & 70--80 & 200 & 280 & 380--390--400 & $\varnothing$ & -- \\
\hline
6 & 3 & 90--100--110 & 210--220 & 290--300--310 & 410 & $\varnothing$ & -- \\
\hline
7 & 4 & 120--130--140 & 230--240--250--260 & 320--330 & 420--430--440 & $\varnothing$ & -- \\
\hline
8 & 4 & 150--160--170--180 & 270--280 & 340 & 450 & $\varnothing$ & -- \\
\hline
9 & 4 & 190--200 & 290--300 & 350--360--370 & 460--470--480--490 & $\varnothing$ & -- \\
\hline
10 & 4 & 210--220 & 310--320 & 380--390--400 & 500--510--520 & $\varnothing$ & -- \\
\hline
\multicolumn{8}{c}{} \\
\hline
\multicolumn{8}{|c|}{Generation 2} \\
\hline
$i$ & $x(i)$ & $\wh{g}_1(i)$ & $\wh{g}_2(i)$ & $\wh{g}_3(i)$ & $\wh{g}_4(i)$ & Par. & Pat. \\
\hline
11 & 2 & 10--40 & 110--130 & 210--220 & 320--330 & $(1,2)$ & (P$_{\ref{Sch22}}$) \\
\hline
12 & 2 & 40--50 & 140--150 & 220--250 & 330--340 & $(2,3)$ & (P$_{\ref{Sch22}}$) \\
\hline
13 & 2 & 20--40 & 110--130 & 210--220 & 310--330 & $(1,2)$ & (P$_{\ref{Sch22}}$) \\
\hline
14 & 3 & 50--60 & 160--170--180 & 250--270 & 340--350--370 & $(3,4)$ & (P$_{\ref{Sch23}}$) \\
\hline
15 & 3 & 40--100--110 & 140--210 & 220--290--310 & 330--410 & $(2,6)$ & (P$_{\ref{Sch23}}$) \\
\hline
16 & 2 & 20--80 & 110--200 & 210--280 & 320--400 & $(1,5)$ & (P$_{\ref{Sch23}}$) \\
\hline
17 & 3 & 50--210--220 & 150--320 & 240--380--400 & 340--520 & $(3,10)$ & (P$_{\ref{Sch24}}$) \\
\hline
18 & 4 & 130--160--180 & 240--250--270 & 320--330--340 & 430--450 & $(7,8)$ & (P$_{\ref{Sch44}}$) \\
\hline
\multicolumn{8}{c}{} \\
\hline
\multicolumn{8}{|c|}{Generation 3} \\
\hline
$i$ & $x(i)$ & $\wh{g}_1(i)$ & $\wh{g}_2(i)$ & $\wh{g}_3(i)$ & $\wh{g}_4(i)$ & Par. & Pat. \\
\hline
19 & 2 & 20--60 & 110--180 & 270--280 & 350--400 & $(4,16)$ & (P$_{\ref{Sch23}}$) \\
\hline
20 & 2 & 40 & 110--150 & 220 & 330 & $(11,12)$ & (P$_{\ref{Sch22}}$) \\
\hline
21 & 4 & 130--180--200 & 270--290--300 & 340--350--370 & 450--480--490 & $(9,18)$ & (P$_{\ref{Sch44}}$) \\
\hline
22 & 3 & 60--210 & 180--320 & 250--390--400 & 370--520 & $(10,14)$ & (P$_{\ref{Sch34}}$) \\
\hline
23 & 3 & 90--130--140 & 220--230--240 & 300--320 & 410--420--440 & $(6,7)$ & (P$_{\ref{Sch34}}$) \\
\hline
24 & 3 & 10--130--160 & 130--270 & 210--330--340 & 330--430--450 & $(11,18)$ & (P$_{\ref{Sch24}}$) \\
\hline
25 & 4 & 190--200 & 290--300 & 350--360--370 & 410--520 & $\varnothing$ & -- \\
\hline
26 & 4 & 130--160--180 & 240--250--270 & 320--330--340 & 410 & $\varnothing$ & -- \\
\hline
\end{tabular}
\medskip
\normalsize
\caption{Full description of generations 1, 2 and 3 in the simulated population.}
\label{TabSimPop12}
\end{table}

\medskip

\begin{table}[h!]
\tiny
\begin{tabular}{|c|c|>{\centering}m{2.3cm}|>{\centering}m{2.3cm}|>{\centering}m{2.3cm}|>{\centering}m{2.3cm}|>{\centering}m{0.9cm}|c|}
\hline
\multicolumn{8}{|c|}{Generation 4} \\
\hline
$i$ & $x(i)$ & $\wh{g}_1(i)$ & $\wh{g}_2(i)$ & $\wh{g}_3(i)$ & $\wh{g}_4(i)$ & Par. & Pat. \\
\hline
27 & 4 & 80--200 & 200--270--300 & 280--340 & 380--400--480--490 & $(5,21)$ & (P$_{\ref{Sch34}}$) \\
\hline
28 & 4 & 40--100--200 & 210--290 & 220--310--350--360 & 330--410--470--490 & $(9,15)$ & (P$_{\ref{Sch34}}$) \\
\hline
29 & 3 & 100--200 & 140--290 & 310--350--370 & 410--460--490 & $(9,15)$ & (P$_{\ref{Sch34}}$) \\
\hline
30 & 4 & 130--200 & 270 & 330--340--350 & 450 & $(18,21)$ & (P$_{\ref{Sch44}}$) \\
\hline
31 & 2 & 20--210 & 180--320 & 280--380 & 400--520 & $(17,19)$ & (P$_{\ref{Sch23}}$) \\
\hline
32 & 2 & 40--60 & 110 & 220--270 & 330--350 & $(19,20)$ & (P$_{\ref{Sch22}}$) \\
\hline
33 & 4 & 10--180--200 & 130--270--380 & 210--340--370 & 430--450--520 & $\varnothing$ & -- \\
\hline
34 & 4 & 20--90--200 & 160--270--330 & 370 & 520--530--550 & $\varnothing$ & -- \\
\hline
35 & 4 & 130--180--200 & 270--290--300 & 340--350--370 & 410 & $(25,26)$ & (P$_{\ref{Sch44}}$) \\
\hline
\multicolumn{8}{c}{} \\
\hline
\multicolumn{8}{|c|}{Generation 5} \\
\hline
$i$ & $x(i)$ & $\wh{g}_1(i)$ & $\wh{g}_2(i)$ & $\wh{g}_3(i)$ & $\wh{g}_4(i)$ & Par. & Pat. \\
\hline
36 & 2 & 60--100 & 180--290 & 270--370 & 340--490 & $(14,29)$ & (P$_{\ref{Sch33}}$) \\
\hline
37 & 3 & 50--200 & 140--160--180 & 250--270--370 & 350--370--410 & $(14,29)$ & (P$_{\ref{Sch33}}$) \\
\hline
38 & 4 & 50--60--100 & 160--170--290 & 270--310--350 & 340--370--410--490 & $(14,29)$ & (P$_{\ref{Sch33}}$) \\
\hline
39 & 2 & 20--210 & 110--150 & 210--380 & 320--340 & $(1,17)$ & (P$_{\ref{Sch23}}$) \\
\hline
40 & 3 & 40--200 & 130--290 & 210--310--370 & 330--410--460 & $(11,29)$ & (P$_{\ref{Sch23}}$) \\
\hline
41 & 2 & 20--110 & 150--320 & 260 & 410--520 & $\varnothing$ & -- \\
\hline
42 & 3 & 230 & 170--390--420 & 240--340 & 380--390 & $\varnothing$ & -- \\
\hline
43 & 4 & 70--90--100 & 210--220--270 & 310--330--340--400 & 490 & $\varnothing$ & -- \\
\hline
\end{tabular}
\medskip
\normalsize
\caption{Full description of generations 4 and 5 in the simulated population.}
\label{TabSimPop345}
\end{table}

\medskip

\begin{table}[h!]
\tiny
\begin{tabular}{|c|c|>{\centering}m{2.3cm}|>{\centering}m{2.3cm}|>{\centering}m{2.3cm}|>{\centering}m{2.3cm}|>{\centering}m{0.9cm}|c|}
\hline
\multicolumn{8}{|c|}{Generation 6} \\
\hline
$i$ & $x(i)$ & $\wh{g}_1(i)$ & $\wh{g}_2(i)$ & $\wh{g}_3(i)$ & $\wh{g}_4(i)$ & Par. & Pat. \\
\hline
44 & 3 & 110--230 & 170--320--390 & 240--260--340 & 390--520 & $(41,42)$ & (P$_{\ref{Sch23}}$) \\
\hline
45 & 2 & 110--230 & 320--420 & 260--340 & 390--520 & $(41,42)$ & (P$_{\ref{Sch23}}$) \\
\hline
46 & 4 & 130--150--160 & 270 & 330--340 & 450 & $(8,18)$ & (P$_{\ref{Sch44}}$) \\
\hline
47 & 2 & 90 & 220 & 310--320 & 410--510 & $\varnothing$ & -- \\
\hline
48 & 3 & 50--210 & 180--320 & 250--400 & 410--520 & $(22,37)$ & (P$_{\ref{Sch33}}$) \\
\hline
49 & 2 & 240 & 320 & 410 & 510--520 & $\varnothing$ & -- \\
\hline
50 & 4 & 100--200 & 270 & 310--330--370 & 410--490--520 & $\varnothing$ & -- \\
\hline
\multicolumn{8}{c}{} \\
\hline
\multicolumn{8}{|c|}{Generation 7} \\
\hline
$i$ & $x(i)$ & $\wh{g}_1(i)$ & $\wh{g}_2(i)$ & $\wh{g}_3(i)$ & $\wh{g}_4(i)$ & Par. & Pat. \\
\hline
51 & 3 & 200--240 & 270--320 & 310--370--410 & 410--490--520 & $(49,50)$ & (P$_{\ref{Sch24}}$) \\
\hline
52 & 4 & 230 & 170--390--420 & 240--340 & 390 & $(42,44)$ & (P$_{\ref{Sch33}}$) \\
\hline
53 & 3 & 130 & 230--240 & 320 & 410--420--440 & $(7,23)$ & (P$_{\ref{Sch34}}$) \\
\hline
\multicolumn{8}{c}{} \\
\hline
\multicolumn{8}{|c|}{Generation 8} \\
\hline
$i$ & $x(i)$ & $\wh{g}_1(i)$ & $\wh{g}_2(i)$ & $\wh{g}_3(i)$ & $\wh{g}_4(i)$ & Par. & Pat. \\
\hline
54 & 4 & 230 & 170--390--420 & 240--340 & 390 & $(42,52)$ & (P$_{\ref{Sch34}}$) \\
\hline
\end{tabular}
\medskip
\normalsize
\caption{Full description of generations 6, 7 and 8 in the simulated population.}
\label{TabSimPop678}
\end{table}

\newpage

\bibliographystyle{acm}
\nocite{*}
\bibliography{PolyGen}

\end{document}